\documentclass{elsarticle}

\usepackage[margin=1.35in]{geometry}
\usepackage{amsmath, amsfonts, amssymb}
\usepackage{hyperref}
\usepackage{cleveref}
\usepackage{caption, subcaption}
\usepackage[frozencache,cachedir=.]{minted}
\usepackage{xcolor}
\usepackage{multirow}

\usepackage{graphicx}

\newtheorem{remark}{Remark}

\newcommand{\jm}[1]{\textcolor{black}{#1}}

\begin{document}
\title{Forward and Inverse Simulation of Pseudo-Two-Dimensional Model of Lithium-Ion Batteries Using Neural Networks}

\author[1]{Myeong-Su Lee \corref{cor1}}
\ead{msl3573@snu.ac.kr}
%\ead[url]{sites.google.com/view/mslee2}

\author[3]{Jaemin Oh \corref{cor1}}
\ead{jaemin_oh@tamu.edu}
%\ead[url]{jaeminoh.github.io}

\author[2]{Dong-Chan Lee}
\author[2]{KangWook Lee}
\author[2]{Sooncheol Park}

\author[1]{Youngjoon Hong \corref{cor2}}
\ead{hongyj@snu.ac.kr}
%\ead[url]{youngjoonhong.com}

\cortext[cor1]{Equal contribution}
\cortext[cor2]{Corresponding author}
\affiliation[1]{organization={Seoul National University}, addressline={1 Gwanak-ro, Gwanak-gu}, postcode={08826}, city={Daejeon}, country={Republic of Korea}}
\affiliation[3]{organization={Texas A\&M University}, addressline={3128 TAMU}, postcode={77843}, city={College Station, TX}, country={USA}}
\affiliation[2]{Organization={Performance Concept Technology Research Lab, Hyundai Motor Company}, addressline={150 Hyundaiyeonguso-Ro, Namyang-Eup, Hwaseong-Si}, postcode={18280}, city={Gyeonggi}, country={Republic of Korea}}

\begin{abstract}
In this work, we address the challenges posed by the high nonlinearity of the Butler--Volmer (BV) equation in forward and inverse simulations of the pseudo-two-dimensional (P2D) model using the physics-informed neural network (PINN) framework.
The BV equation presents significant challenges for PINNs, primarily due to the hyperbolic sine term, which renders the Hessian of the PINN loss function highly ill-conditioned. To address this issue, we introduce a bypassing term that improves numerical stability by substantially reducing the condition number of the Hessian matrix. 
Furthermore, the small magnitude of the ionic flux \( j \) often leads to a common failure mode where PINNs converge to incorrect solutions. We demonstrate that incorporating a secondary conservation law for the solid-phase potential \( \psi \) effectively prevents such convergence issues and ensures solution accuracy. 
The proposed methods prove effective for solving both forward and inverse problems involving the BV equation. Specifically, we achieve precise parameter estimation in inverse scenarios and reliable solution predictions for forward simulations.
\end{abstract}


\maketitle

\section{Introduction}\label{sec-introduction}
Lithium-ion (Li-ion) batteries have become integral to modern technology, powering a wide range of devices from laptops and mobile phones to electric vehicles \cite{zubi2018lithium}. The 2019 Nobel Prize in Chemistry, awarded to the pioneers of Li-ion battery technology, highlights its transformative impact on modern life. However, despite their widespread adoption, Li-ion batteries face inherent challenges, including finite lifecycles, performance degradation, and safety risks such as thermal runaway and explosions \cite{bandhauer2011critical, etacheri2011challenges}.
These limitations, alongside environmental and economic concerns, underscore the critical need for precise monitoring and effective management of battery systems. Ensuring reliability and safety while maximizing performance and lifespan is not only vital for current applications but also for advancing the next generation of energy storage technologies \cite{costa2021recycling}.

Effective battery management relies heavily on the ability to accurately predict and understand changes in the battery state under varying operating and environmental conditions \cite{xiong2018towards, wang2020comprehensive}. In this context, physics-based model simulations have emerged as a powerful tool, enabling accurate battery state estimation without the need for extensive experimental data. These simulations play a pivotal role in solving both forward problems, such as predicting battery performance, and inverse problems, such as identifying key material properties and reaction kinetics from limited data \cite{nikdel2014various, plett2004extended}. Moreover, the increasing emphasis on digital twin technologies highlights the central role of physics-based models in developing intelligent and efficient battery management systems. By enabling a deeper understanding of battery behavior, these models support not only safer and more sustainable energy storage solutions but also the optimization of battery design and functionality \cite{li2020digital}.
The pseudo-two-dimensional (P2D) model, also known as the Doyle--Fuller--Newman model, has been widely used in Li-ion battery studies due to its ability to accurately capture the essential electrochemical processes, such as transport and reaction kinetics \cite{doyle1993modeling}. The P2D model divides the battery into three distinct regions: the negative electrode, the separator, and the positive electrode. It accounts for the transport of lithium ions through the electrolyte, the diffusion of lithium within solid particles in the electrodes, and the electrochemical reactions that occur at the interfaces between the electrolyte and solid particles. The P2D model consists of coupled partial differential equations (PDEs) to describe the concentration of lithium ions and the electric potentials across these regions, along with the Butler--Volmer (BV) equation to model the kinetics of the electrochemical reactions. Numerous studies have validated the P2D model's accuracy, showing good agreement with experimental data and highlighting its effectiveness in predicting the dynamic behavior of Li-ion batteries \cite{doyle1996comparison,kumaresan2007thermal}.

Numerical experiments play a crucial role in understanding and optimizing complex systems, including Li-ion battery models. Over the years, numerous efforts have been made to develop and refine classical numerical methods for solving PDEs, such as finite element and finite difference methods. These methods have proven effective in solving the P2D model, providing accurate results under various conditions \cite{cai2011mathematical, han2021fast, mora2024high}. The availability of software packages in MATLAB \cite{torchio2016lionsimba}, Python \cite{sulzer2021python}, and Julia \cite{berliner2021petlion} has further democratized the use of these methods, making them accessible to researchers and engineers across disciplines. However, the complexity of the underlying equations presents implementation challenges, and these classical methods often struggle with solving inverse problems due to high computational demands and sensitivity to noise.
Recently, scientific machine learning (SciML) has emerged as a potential alternative method for solving both forward and inverse problems of PDEs. Especially, physics-informed neural networks (PINNs) incorporate governing physical laws directly into the loss function used for neural network learning, making them particularly well-suited for problems with limited data \cite{raissi2019physics}. PINNs have shown promising results across various scientific domains \cite{he2020physics, jo2024density, son2024pinn}. These applications highlight PINNs' ability to handle complex, multi-scale, and multi-physics problems by leveraging the underlying physics as a constraint during the optimization process.
Given this success in other fields, there has been increasing interest in applying SciML techniques, particularly PINNs, to battery simulations. In the context of battery simulations, SciML techniques hold significant potential for capturing hidden physical phenomena, such as transport and reaction dynamics, while also enabling the estimation of critical parameters that are difficult to measure directly.

Despite the notable progress in applying PINNs to various fields of physics and engineering, their applicability across diverse domains still remains underexplored, highlighting the need for further extensive research. In particular, vanilla PINNs have shown significant limitations in addressing highly non-linear phenomena, complex boundary conditions, and stiff problems. These challenges also arise in battery simulations \cite{ostanek2023novel}. Specifically, the vanilla PINN approach struggles to effectively solve the P2D model due to the highly nonlinear flux term, known as the BV equation, which describes electrochemical reactions in batteries, as well as the model’s multi-scale nature. To address these challenges, several studies have proposed modified approaches. Xue et al. \cite{xue2023enhanced} considered the single particle model (SPM), a simplified version of the P2D model, instead of directly solving the full P2D model, to reduce computational complexity and instability. Hassanaly et al. \cite{hassanaly2023pinn2} proposed a two-step training strategy: first fitting the neural network to the single particle model (SPM), and then using the learned parameters as initial values for fitting the P2D model with a linearized BV equation. For the nonlinear BV equation, they suggested an additional division step to handle gradient exploding problems. Besides, Zubov et al. \cite{zubov2021neuralpde} studied a simplified P2D model using PINNs to test their software \texttt{NeuralPDE.jl}. Chen et al. \cite{chen2023physics} simulated redox flow batteries with PINNs, formulating a model with 6 governing equations and 24 boundary conditions.

In this work, we address the specific challenges posed by the BV equation in P2D model simulations using PINNs. By designing and employing simplified toy models that replicate the inherent complexities of the BV equation, we systematically analyzed and identified critical issues that arise in battery simulations. Based on these insights, we developed novel strategies to overcome these challenges, enabling the direct solution of the full P2D model with the fully nonlinear BV equation in a single step, without relying on commonly used simplifications such as the SPM or linearized BV equations. 
The proposed strategies include the following. First, we introduce a bypassing term, which employs a neural network to approximate the stiffest component of the BV equation. This reduces the Hessian spectrum of the PINN loss, significantly improving the conditioning of the optimization problem and stabilizing the training process. Second, we augment the PINN loss function with a secondary conservation law for the solid-phase potential, ensuring that the predicted solution faithfully captures the nonlinear dynamics of the system and avoids convergence to incorrect solutions. 
Using these strategies, we achieve accurate simulation results. This demonstrates the efficacy of our approach in addressing the computational challenges associated with nonlinear battery models.
Furthermore, we demonstrate the effectiveness of the proposed PINN framework in addressing inverse problems related to Li-ion battery models, specifically the estimation of critical geometric parameters such as the total battery length and the length ratio of battery sections. Inverse simulations of this kind are notoriously challenging for classical numerical methods due to high computational costs and sensitivity to noise. By leveraging the strengths of the PINN framework and building on our robust solution to the forward problem, we successfully estimate these parameters directly within the nonlinear P2D model framework. Our method requires minimal modifications to incorporate observational data and avoids the need for simplifying assumptions or additional numerical approximations. These results highlight the capability of our approach to tackle complex inverse problems with improved reliability and computational efficiency.
\\

\section{Preliminary}
We start with a brief overview of the P2D model, followed by an introduction to the basic concepts of PINNs.

\subsection{Pseudo-two-dimensional model}\label{sec-p2d} 
% Lithium-ion battery cells include three primary components: positive electrode, separator, and negative electrode. Both electrodes contain solid particles. The electrolyte, which permeates both the separator and the porous structure of the electrodes, serves as the medium through which lithium ions move between the electrodes. When the battery is charged or discharged, lithium ions migrate through the electrolyte from the negative electrode to the positive electrode, or vice versa, depending on the direction of the current. During this process, lithium ions enter and exit the solid particles in the electrodes, causing changes in the concentration of lithium within the particles. These changes affect the electrode's potential and influence the overall electrochemical reactions at the interface. This dynamic process of Li-ion in battery cells plays a critical role in determining the battery's capacity, efficiency, and performance.
Lithium-ion battery cells consist of three main components: the positive electrode, separator, and negative electrode. Both the positive and negative electrodes contain solid particles, and the electrolyte permeates the porous structure of these electrodes as well as the separator, allowing lithium ions to move between the electrodes. When the battery is charged or discharged, lithium ions migrate through the electrolyte from the negative electrode to the positive electrode, or vice versa, driven by the potential difference between the electrodes. During this process, lithium ions enter or exit the solid particles within the electrodes, causing changes in lithium concentration inside these particles as a result of electrochemical reactions occurring at the interfaces. These electrochemical reactions and resulting concentration changes play a crucial role in determining the battery’s capacity, efficiency, and performance.

\begin{figure}[t!]
    \centering
    \includegraphics[width=0.7\linewidth]{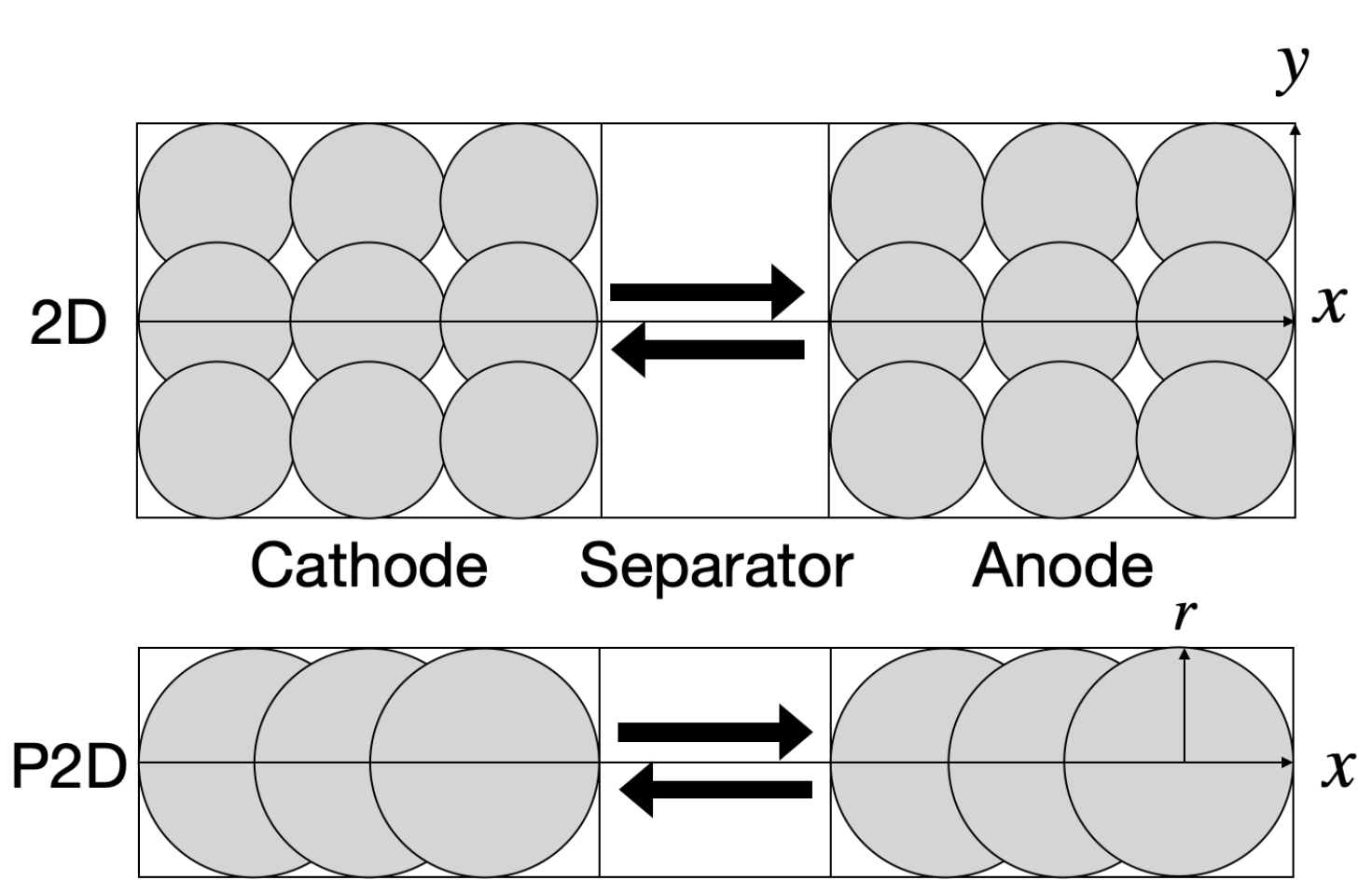}
    \caption{\textcolor{black}{A schematic of the P2D model.}}
    \label{fig-geometry_p2d}
\end{figure}

The P2D model, proposed in \cite{doyle1993modeling}, captures the complex dynamics within a Li-ion battery cell by modeling the transport and diffusion of lithium ions along the through-thickness of the cell, denoted as the $x$-direction. The model assumes that lithium ions move through the electrolyte in the $x$-direction, while Li-ion diffusion within each spherical solid particle in the electrodes occurs symmetrically along the radial direction $r$. In this paper, we denote the lengths of the positive electrode, separator, and negative electrode as $L_p$, $L_s$, and $L_n$, respectively. Then, $x \in [0, L_p] =: \mathcal{D}_p$ corresponds to the positive electrode, $x \in [L_p, L_p + L_s] =: \mathcal{D}_s$ represents the separator, and $x \in [L_p + L_s, L_p + L_s + L_n] =: \mathcal{D}_n$ represents the negative electrode. For simplicity, we denote the total length as $L:= L_p + L_s + L_n$. For the radial direction within solid particles, the coordinate $r$ ranges from $0$ to $\mathcal{R}_p$ in the positive electrode and from $0$ to $\mathcal{R}_n$ in the negative electrode, where $\mathcal{R}_p$ and $\mathcal{R}_n$ denote the radii of the solid particles in each electrode, respectively. \textcolor{black}{For descriptive purposes, we provide a schematic representation of the P2D model in \Cref{fig-geometry_p2d}.}

The primary variables of the P2D model include: (1) the solid-phase Li-ion concentration, \( c_{s,i} \) (\( i = p, n \)), in the positive and negative electrodes; (2) the solid-phase potential, \( \psi_i \) (\( i = p, n \)), in the electrodes; (3) the Li-ion concentration, \( c_i \) (\( i = p, s, n \)), in the electrolyte; and (4) the liquid-phase potential, \( \phi_i \) (\( i = p, s, n \)), in the electrolyte. Here, \( p \), \( s \), and \( n \) represent the positive electrode, separator, and negative electrode, respectively. Additional variables, such as the overpotential \(\eta_i\) and ionic flux \(j_i\), are also considered. A comprehensive summary of these state variables and other quantities used in the P2D model is provided in \Cref{tab-variables}. Using these variables, the governing equations of the P2D model are formulated to describe the time evolution of the state variables as follows.

\noindent
{\bf Solid phase concentration} (Fick's Second Law in Spherical Coordinates): For each $i=p,n$,
\begin{equation*}\label{eq-cs-pde}
    \dfrac{\partial c_{s,i}}{\partial t} = \dfrac{1}{r^2}\dfrac{\partial}{\partial r}\left( r^2 D_{s,i} \dfrac{\partial c_{s,i}}{\partial r} \right), \qquad r\in(0,\mathcal{R}_i).
\end{equation*}
The boundary conditions are specified as follows:
%\begin{equation*}
\[
\left\{ \begin{aligned} 
    -D_{s,i}^\mathrm{eff} \dfrac{\partial c_{s,i}}{\partial r} = 0, \quad &\text{at } r = 0,\\ 
    -D_{s,i}^\mathrm{eff} \dfrac{\partial c_{s,i}}{\partial r} = j_i, \ 
    &\text{at}\ r=\mathcal{R}_i.
\end{aligned} \right.
%\end{equation*}
\]
{\bf Liquid phase concentration}: For $i=p, s, n$, we write
\begin{equation*}
    \epsilon_i \dfrac{\partial c_i}{\partial t} = \dfrac{\partial }{\partial x} \left(D_i^\mathrm{eff} \dfrac{\partial c_i}{\partial x}\right) + (1 - t_+^0)a_i j_i.
\end{equation*}
The boundary conditions are given as follows:
\begin{align*}
    -D_p^\mathrm{eff} \dfrac{\partial c_p}{\partial x} = 0\ \text{at}\ x=0\qquad \text{and}\qquad -D_n^\mathrm{eff} \dfrac{\partial c_n}{\partial x} = 0\ \text{at}\ x=L.
\end{align*}
The interfacial conditions are given to impose continuity and their flux continuity at the interfaces:
\begin{align*}
\left\{
    \begin{array}{l}
        c_p = c_s \\
        -D_p^\mathrm{eff} \dfrac{\partial c_p}{\partial x} = -D_s^\mathrm{eff} \dfrac{\partial c_s}{\partial x}
    \end{array}
\right. &\qquad x = L_p \qquad \text{and} \qquad
\left\{
    \begin{array}{l}
        c_s = c_n \\
        -D_s^\mathrm{eff} \dfrac{\partial c_s}{\partial x} = -D_n^\mathrm{eff} \dfrac{\partial c_n}{\partial x}
    \end{array}
\right. &\qquad x = L_p + L_s.
\end{align*}
{\bf Solid Phase Potential} (Ohm's Law): For each $i=p,n$,
\begin{equation*}\label{eq-psi-pde}
\left\{
    \begin{array}{ll}
        \sigma_p^\mathrm{eff} \dfrac{\partial^2 \psi_p}{\partial x^2} = a_pFj_p & \qquad x \in (0, L_p) \\[8pt]
        -\sigma_p^\mathrm{eff} \dfrac{\partial \psi_p}{\partial x} = I_\mathrm{app} & \qquad x = 0, \\[8pt]
        -\sigma_p^\mathrm{eff} \dfrac{\partial \psi_p}{\partial x} = 0 & \qquad x = L_p
    \end{array}
\right.
\qquad \text{and} \qquad
\left\{
    \begin{array}{ll}
        \sigma_n^\mathrm{eff} \dfrac{\partial^2 \psi_n}{\partial x^2} = a_nFj_n & \qquad x \in (0, L_n) \\[8pt]
        -\sigma_n^\mathrm{eff} \dfrac{\partial \psi_n}{\partial x} = 0 & \qquad x = L_p + L_s, \\[8pt]
        \psi_n = 0 & \qquad x = L,
    \end{array}
\right.
\end{equation*}
where $I_{\text{app}}$ is the applied current density.

\noindent
{\bf Liquid phase potential} (based on Ohm's Law): For $i=p,s,n$,
\begin{equation*}\label{eq-phi-pde}
    -\dfrac{\partial }{\partial x}\left(\kappa_i^\mathrm{eff} \dfrac{\partial \phi_i}{\partial x} \right) + \dfrac{2RT(1 - t_+^0)}{F}\dfrac{\partial }{\partial x} \left( \kappa_i^\mathrm{eff} \dfrac{\partial \log c_i}{\partial x} \right) = a_iFj_i.
\end{equation*}
The boundary conditions are given as follows:
\begin{align*}
    -\kappa_p^\mathrm{eff} \dfrac{\partial \phi_p}{\partial x}=0\qquad \text{at}\ x=0\qquad \text{and}\qquad
    -\kappa_n^\mathrm{eff} \dfrac{\partial \phi_n}{\partial x}=0\qquad \text{at}\ x=L.
\end{align*}
The interfacial conditions are given to impose continuity and their flux continuity at the interfaces:
\begin{align*}
\left\{
    \begin{array}{l}
        \phi_p = \phi_s \\[8pt]
        -\kappa_p^\mathrm{eff} \dfrac{\partial \phi_p}{\partial x} = -\kappa_s^\mathrm{eff} \dfrac{\partial \phi_s}{\partial x}
    \end{array}
\right. &\qquad x = L_p \qquad \text{and} \qquad
\left\{
    \begin{array}{l}
        \phi_s = \phi_n \\[8pt]
        -\kappa_s^\mathrm{eff} \dfrac{\partial \phi_s}{\partial x} = -\kappa_n^\mathrm{eff} \dfrac{\partial \phi_n}{\partial x}
    \end{array}
\right. &\qquad x = L_p + L_s.
\end{align*}

\noindent
The aforementioned state variables are coupled by the ionic flux $j_i$, which is given by the following BV equation:  
\begin{equation}\label{BV}
    j_i = 
    \begin{cases}
        2 k_i \left(c_{s,i,\mathrm{max}} - c_{s,i,\mathrm{surf}}\right)^{0.5} c_{s,i,\mathrm{surf}}^{0.5} c_i^{0.5} \sinh\left(\dfrac{F}{2RT}\eta_i\right) \qquad &\text{for}\ i=p,n\\
        0&\text{for}\ i=s
    \end{cases}
\end{equation}
where \(c_{s, i, \mathrm{surf}} = c_{s, i}(\mathcal{R}_i)\), and \(\eta_i\) represents the over-potential defined as:
\begin{equation*}
    \eta_i = \psi_i - \phi_i - U_i\left(\dfrac{c_{s,i}}{c_{s,i,\mathrm{max}}}\right),
\end{equation*}
and \(U_i\) denotes the open circuit potential.\\

Interested readers may consult with \cite{diaz2018well} for well-posedness, and \cite{torchio2016lionsimba, cai2011mathematical} for numerics.

\begin{table}[ht]
\centering
% \scriptsize
\begin{tabular}{c|c|c|c}
    \hline\hline
    Quantity & Description & Quantity & Description\\
    \hline
    \(\psi\) & solid phase potential & $R$ & gas constant \\
    \(\phi\) & liquid phase potential & $F$ & Faraday's constant \\
    \(c_s\) & solid phase concentration &  \(c_{s, \mathrm{max}}\) & maximum value for $c_s$ \\
    \(c\) & liquid phase concentration & \(c_\mathrm{ref}\) & reference value for $c$ \\
    $j$ & ionic flux & \(k\) & electrochemical reaction rate constant \\
    $T$ & temperature & $T_\mathrm{ref}$ & Reference temperature \\
    $D$ & diffusion coefficient & $D_s$ & diffusion coefficient for solid particle \\
    $U$ & open circuit potential & \(I_\mathrm{app}\) & applied current density \\
    $a$ & area &  \(\sigma\) & electronic conductivity of solid matrix \\
    $\epsilon$ & volume fraction & \(t_+^0\) & transference number of Li-ion \\
    $\eta$ & overpotential & Brugg & Bruggmann constant \\
    $\kappa$ & ionic conductivity & $\bullet^\mathrm{eff}$ & effective coefficient: $\bullet \times \epsilon^\mathrm{Brugg}$ \\
    \hline\hline
\end{tabular}
\caption{A list of battery state variables and other quantities.}\label{tab-variables}
\end{table}

\subsection{Physics-informed neural networks}\label{sec-pinn}
Physics-Informed Neural Networks (PINNs), as proposed in \cite{raissi2019physics}, are a deep learning framework designed to solve both forward and inverse problems for partial differential equations (PDEs) within a unified framework. 

For simplicity, we consider the following simple example. Let $\Omega \subset \mathbb{R}^{d_\mathrm{in}}$ be a domain and $u: \mathbb{R}^{d_\mathrm{in}} \rightarrow \mathbb{R}^{d_\mathrm{out}}$ be the solution to the following differential equation:
\begin{align*}
    \mathcal{N}[u](x) &= f(x), \quad x \in \Omega, \\
    \mathcal{B}[u](x) &= g(x), \quad x \in \partial \Omega,
\end{align*}
where $\mathcal{N}$ is a general differential operator and $\mathcal{B}$ is a boundary operator. Note that the boundary condition can also represent initial conditions in time-dependent problems. PINNs approach approximates the solution $u$ by using a neural network approximation $u_{\theta^\star}$, where $\theta^\star$ denotes the optimal set of neural network parameters. These parameters are obtained by solving the following minimization problem:
\begin{equation*}
    \theta^\star = \arg \min_\theta \mathcal{L}_\mathrm{PINN}(\theta),
\end{equation*}
where the loss function $\mathcal{L}_\mathrm{PINN}(\theta)$ is defined as:
\begin{equation}\label{eq-pinn-loss}
    \mathcal{L}_\mathrm{PINN}(\theta) = \int_\Omega \left\lvert\mathcal{N}[u_\theta](x) - f(x)\right\rvert_2^2 \mathrm{d}x + \lambda \int_{\partial \Omega} \left\lvert \mathcal{B}[u_\theta](s) - g(s) \right\rvert_2^2\mathrm{d}\sigma(s).
\end{equation}
Here, $\sigma$ is a surface measure, and $\lambda$ is a weighting parameter balancing the importance of the differential equation and the boundary conditions. In practice, the integrals in \eqref{eq-pinn-loss} are approximated using numerical quadrature methods. Common approaches include the trapezoidal rule \cite{trefethen2014exponentially}, Gauss quadrature \cite{golub1969calculation}, and Quasi-Monte Carlo methods, such as the Sobol sequence \cite{sobol1967distribution}. The choice of the quadrature method can significantly impact the accuracy and efficiency of the PINN approach and may depend on the specific problem and domain characteristics.

\section{\textcolor{black}{Conventional application of PINNs to the P2D model}}
\label{sec-pinn-p2d}

\textcolor{black}{
In this section, we first provide a complete presentation of how PINNs are typically applied to PDEs, which we refer to as the ``vanilla PINN'' approach. We then discuss some issues that can appears in applying the vanilla PINNs and address the issues with existing methods.}

\subsection{\textcolor{black}{Vanilla PINNs for the P2D model}}
{\color{black}In this subsection, we present a complete explanation of a vanilla PINN approach to solve the P2D model. The P2D model includes two interfaces: one between the positive electrode and the separator, and the other between the separator and the negative electrode. The interfacial conditions enforce that for the liquid-phase concentration $c_{i}$ and the potential $\psi_i$, they and their flux are continuous at the interfaces of both the electrodes and the separator. The flux continuity conditions can cause discontinuities in the first derivatives, which makes it difficult for a single approximate network defined over the whole $x$-axis to satisfy the flux continuity. To address this, we adopt a \emph{domain decomposition} approach, dividing the $x$-axis into three subdomains $\mathcal{D}_p$, $\mathcal{D}_s$, and $\mathcal{D}_n$. We then assign a separate neural network to approximate each primary state variable (e.g. $c_{s,i}, c_i, \phi_i, \psi_i$) in each subdomain. In other words, each region $i \in \{p, s, n\}$ is handled by its own network, denoted by
\[
\widetilde{c}_{s,i}, \quad \widetilde{c}_{i}, \quad \widetilde{\phi}_i, \quad \widetilde{\psi}_i.
\]
With such approximations for the primary variables by neural networks, the ionic flux $j_i$ is approximated by
\begin{equation}\label{eq-pinn-j}
    \widetilde{j}_i = 2 k \left(c_{s,\mathrm{max}} - \widetilde{c}_{s,i,\mathrm{surf}}\right)^{0.5} \widetilde{c}_{s,i,\mathrm{surf}}^{0.5} \widetilde{c}_i^{0.5} \sinh\left(\frac{F}{2RT}\left(\widetilde{\phi}_i-\widetilde{\psi}_i-U(\widetilde{c}_{s,i}/c_\mathrm{s,max})\right)\right),
\end{equation}
according to the original relation \eqref{BV}. Then, the PINN loss function is given as follows:
\begin{align}\label{eq-pinn-loss2}
    \mathcal{L}_\mathrm{PINN}(\theta) := \mathcal{L}_\mathrm{PDE}(\theta)+\lambda_\mathrm{IC}\mathcal{L}_\mathrm{IC}(\theta)+\lambda_\mathrm{BC}\mathcal{L}_\mathrm{BC}(\theta)+\lambda_\mathrm{Inter}\mathcal{L}_\mathrm{Inter}(\theta),
\end{align}
where $\theta$ represents the network parameter and each of the loss function terms and $\lambda_{\bullet}$ are weighting parameters to be chosen appropriately. Each term in the loss function---$\mathcal{L}_\mathrm{PDE}(\theta)$, $\mathcal{L}_\mathrm{IC}(\theta)$, $\mathcal{L}_\mathrm{BC}(\theta)$, and $\mathcal{L}_\mathrm{Inter}(\theta)$---is conventionally defined to enforce the PDEs, boundary/initial conditions, and the interface conditions into the networks.  

Specifically, the PDE residual loss $\mathcal{L}_\mathrm{PDE}(\theta)$ can be split into four parts, corresponding to the main variables: $\mathcal{L}_\mathrm{c_s, PDE}(\theta)$, $\mathcal{L}_\mathrm{c, PDE}(\theta)$, $\mathcal{L}_\mathrm{\phi, PDE}(\theta)$, and $\mathcal{L}_\mathrm{\psi, PDE}(\theta)$. For example, the PDE residual loss $\mathcal{L}_\mathrm{c,PDE}(\theta)$ for the liquid concentration $c$ is given as
\begin{align*}
    \begin{split}
        \mathcal{L}_\mathrm{c, PDE}(\theta)=\sum_{i\in\{p,s,n\}}\int_{[0,\mathcal{T}]\times\mathcal{D}_i\times[0,\mathcal{R}_i]}\left|
    \epsilon_i\frac{\partial \widetilde{c}_i}{\partial t} - \frac{\partial }{\partial x} \left(D^\mathrm{eff} \frac{\partial \widetilde{c}_i}{\partial x}\right) - (1 - t_+^0)a_i \widetilde{j}_i\right|^2\ \mathrm{d}t\mathrm{d}x\mathrm{d}r,
    \end{split}.
\end{align*}
The PDE residual losses for the other state variables are given similarly. Likewise, the initial, boundary, and interface condition losses can also be split into four parts. For instance, the losses for the liquid-phase concentration $c_i$ are given as follows:
\begin{align*}
    \mathcal{L}_\mathrm{c,IC}(\theta)=&\sum_{i=p,s,n}\int_{\mathcal{D}_i}\left|\widetilde{c}_{i}(0,x)-c_{i}^0(x)\right|^2\mathrm{d}x,\\
    \mathcal{L}_\mathrm{c,BC}(\theta)=&\int_{[0,\mathcal{T}]}\left|D_p^\mathrm{eff}\frac{\partial\widetilde{c}_{p}}{\partial x}(t,0)\right|^2+\left|D_n^\mathrm{eff}\frac{\partial\widetilde{c}_{n}}{\partial x}(t,L)\right|^2\mathrm{d}t,\\
    \mathcal{L}_\mathrm{c,Inter}(\theta)=&\int_{[0,\mathcal{T}]}\left|D_p^\mathrm{eff}\frac{\partial\widetilde{c}_{p}}{\partial x}(t,L_p)-D_s^\mathrm{eff}\frac{\partial\widetilde{c}_{s}}{\partial x}(t,L_p)\right|^2
    +\left|\widetilde{c}_p(t,L_p)-\widetilde{c}_s(t,L_p)\right|^2\\
    +&\left|D_s^\mathrm{eff}\frac{\partial\widetilde{c}_{s}}{\partial x}(t,L_p+L_s)-D_n^\mathrm{eff}\frac{\partial\widetilde{c}_{n}}{\partial x}(t,L_p+L_s)\right|^2+\left|\widetilde{c}_s(t,L_p+L_s)-\widetilde{c}_n(t,L_p+L_s)\right|^2\mathrm{d}t.
\end{align*}
The boundary, initial, and interface losses for other variables are defined similarly, following the same manners. 
While the vanilla PINN approach, in principle, be applied to a wide range of PDEs, PINNs have often been shown to exhibit failure modes \cite{krishnapriyan2021characterizing}, particularly for strongly nonlinear PDE systems. Indeed, when applying this straightforward method to the P2D model, one encounters such failure modes due to inherent complexities. In the remainder of this section, we discuss several  challenges with the vanilla approach and address them by using standard strategies.
}
\textcolor{black}{\subsection{Issues with vanilla PINNs}
This subsection is devoted to addressing several issues that arise when applying the above vanilla approach to solve the P2D model.}
\noindent\subsubsection*{Multi-scale issues}

One of the challenges in applying PINNs to the P2D model lies in the significant scale differences between the microscopic and macroscopic coordinates, as well as the various physical and chemical constants in the system. The P2D model considers two spatial coordinates: (1) macroscopic coordinates $x$, representing the through-thickness position across the electrolyte and electrodes, and (2) microscopic coordinates $r$, representing the radial position within solid particles.
These scale differences pose numerical difficulties in the optimization process, particularly when the scales differ by several orders of magnitude. Additionally, the physical and chemical constants, such as diffusion coefficients and conductivity, exhibit vast disparities in scale, further complicating the learning process in the PINN framework.

Such scale-related issues have been widely reported not only in classical numerical methods but also in several SciML methods. One of the simplest yet effective techniques to address these issues is non-dimensionalization. Following this, we adopt a non-dimensionalized form of the P2D model to mitigate the multi-scale issues. Specifically, the coordinate components $t$, $x$, and $r$, and the state variables $c_{s,i}$, $c_i$, $\phi_i$, and $\psi_i$ are rescaled as follows:
\begin{align*}
    \hat{t}=\frac{t}{\mathcal{T}},\quad \hat{x}=\frac{x}{L},\quad \hat{r}=\frac{r}{\mathcal{R}_i},\quad \hat{c}_{s,i}=\frac{c_{s,i}}{c_{s,\mathrm{max}}},\quad \hat{c}_{i}=\frac{c_i}{c_\mathrm{ref}},\quad \hat{\phi_i}=\frac{\phi_i}{\phi_\mathrm{ref}},\quad \hat{\psi}_i=\frac{\psi_i}{\psi_\mathrm{ref}},
\end{align*}
which yields $\hat{t},\ \hat{x},\ \hat{r} \in [0,1]$, $\hat{c}_{s,i} \in [0,1]$, and $\hat{c}_i$, $\hat{\phi}_i$, $\hat{\psi}_i \sim \mathcal{O}(1)$. Following this non-dimensionalization, the rescaled governing equations are given by:

\noindent
{\bf Solid phase concentration}: For $i=p,s,n$,
\begin{equation*}\label{eq-cs-pde-nd}
    \hat{r}\dfrac{\partial \hat{c}_{s,i}}{\partial \hat{t}} = \dfrac{\mathcal{T}D_{s,i}}{R_{i}^2}\left(2\dfrac{\partial \hat{c}_{s,i}}{\partial\hat{r}}+\dfrac{\partial^2\hat{c}_{s,i}}{\partial\hat{r}^2}\right).
\end{equation*}
The boundary conditions are specified as follows:
\begin{equation*}
\left\{
    \begin{array}{ll}
        \dfrac{\partial \hat{c}_{s,i}}{\partial \hat{r}} = 0 & \quad \text{at } \hat{r} = 0, \\[8pt]
        \dfrac{\partial \hat{c}_{s,i}}{\partial \hat{r}} = \dfrac{\mathcal{R}_i j_i}{-D_{s,i}^\mathrm{eff}} & \quad \text{at } \hat{r} = 1,
    \end{array}
\right.
\end{equation*}
{\bf Liquid phase concentration}: For each $i=p,s,n$,
\begin{equation*}
    \frac{\partial \hat{c}_i}{\partial \hat{t}} = \frac{D_i^\mathrm{eff}\mathcal{T}}{L^2\epsilon_i}\frac{\partial^2\hat{c_i} }{\partial \hat{x}^2} + \frac{(1 - t_+^0)\mathcal{T}a_i j_i}{\epsilon_ic_{\text{ref}}}.
\end{equation*}
The boundary conditions are given as follows:
\begin{align*}
    \dfrac{\partial \hat{c}_p}{\partial \hat{x}} = 0\qquad \text{at}\ \hat{x}=0\qquad \text{and}\qquad  \dfrac{\partial \hat{c}_n}{\partial \hat{x}} = 0\qquad \text{at}\ \hat{x}=1.
\end{align*}
The interfacial conditions are given to impose continuity and their flux continuity at the interfaces:
\begin{align*}
\left\{
    \begin{array}{l}
        \hat{c}_p = \hat{c}_s \\[8pt]
        -D_p^\mathrm{eff} \dfrac{\partial \hat{c}_p}{\partial \hat{x}} = -D_s^\mathrm{eff} \dfrac{\partial \hat{c}_s}{\partial \hat{x}}
    \end{array}
\right. &\qquad \hat{x} = \dfrac{L_p}{L} \qquad \text{and} \qquad
\left\{
    \begin{array}{l}
        c_s = c_n \\[8pt]
        -D_s^\mathrm{eff} \dfrac{\partial c_s}{\partial x} = -D_n^\mathrm{eff} \dfrac{\partial c_n}{\partial x}
    \end{array}
\right. &\qquad \hat{x} = \dfrac{L_p + L_s}{L}.
\end{align*}
{\bf Solid phase potential}: For $i=p,n$,
\begin{equation*}\label{eq-psi-pde-nd}
\left\{
    \begin{array}{ll}
        \dfrac{\partial^2 \hat{\psi}_p}{\partial \hat{x}^2} = \dfrac{L^2 a_p F j_p}{\sigma_p^\mathrm{eff} \psi_{\text{ref}}} & \qquad \hat{x} \in (0, L_p / L) \\[8pt]
        \dfrac{\partial \hat{\psi}_p}{\partial \hat{x}} = \dfrac{L^2 I_\mathrm{app}}{-\sigma_p^\mathrm{eff} \psi_{\text{ref}}} & \qquad \hat{x} = 0, \\[8pt]
        \dfrac{\partial \hat{\psi}_p}{\partial \hat{x}} = 0 & \qquad \hat{x} = L_p / L
    \end{array}
\right.
\qquad \text{and} \qquad
\left\{
    \begin{array}{ll}
        \dfrac{\partial^2 \hat{\psi}_n}{\partial x^2} = \dfrac{L^2 a_n F j_n}{\sigma_n^\mathrm{eff} \psi_{\text{ref}}} & \qquad \hat{x} \in (0, L_n / L) \\[8pt]
        \dfrac{\partial \hat{\psi}_n}{\partial \hat{x}} = 0 & \qquad \hat{x} = (L_p + L_s) / L, \\[8pt]
        \hat{\psi}_n = 0 & \qquad \hat{x} = 1.
    \end{array}
\right.
\end{equation*}

\noindent
{\bf Liquid phase potential}: For $i=p,s,n$,
\begin{equation*}\label{eq-phi-pde-nd}
    -\dfrac{\partial }{\partial \hat{x}}\left(\kappa_i^\mathrm{eff} \dfrac{\partial \hat{\phi}_i}{\partial \hat{x}} \right) + \dfrac{2RT(1 - t_+^0)}{F}\dfrac{\partial }{\partial \hat{x}} \left( \kappa_i^\mathrm{eff} \dfrac{\partial \log \hat{c}_i}{\partial \hat{x}} \right) = L^2 a_iFj_i.
\end{equation*}
The boundary conditions are given as follows:
\begin{align*}
    -\kappa_p^\mathrm{eff} \dfrac{\partial \hat{\phi}_p}{\partial \hat{x}}=0\qquad \text{at}\ \hat{x}=0\qquad \text{and}\qquad
    -\kappa_n^\mathrm{eff} \dfrac{\partial \hat{\phi}_n}{\partial \hat{x}}=0\qquad \text{at}\ \hat{x}=1.
\end{align*} 
The interfacial conditions are given as follows:
\begin{align*}
\left\{
    \begin{array}{l}
        \hat{\phi}_p = \hat{\phi}_s \\[8pt]
        -\kappa_p^\mathrm{eff} \dfrac{\partial \hat{\phi}_p}{\partial \hat{x}} = -\kappa_s^\mathrm{eff} \dfrac{\partial \hat{\phi}_s}{\partial \hat{x}}
    \end{array}
\right. &\qquad \hat{x} = \dfrac{L_p}{L} \qquad \text{and} \qquad
\left\{
    \begin{array}{l}
        \hat{\phi}_s = \hat{\phi}_n \\[8pt]
        -\kappa_s^\mathrm{eff} \dfrac{\partial \hat{\phi}_s}{\partial \hat{x}} = -\kappa_n^\mathrm{eff} \dfrac{\partial \hat{\phi}_n}{\partial \hat{x}}
    \end{array}
\right. &\qquad \hat{x} = \dfrac{L_p + L_s}{L}.
\end{align*}

\noindent
{\bf Butler--Volmer equation} With the rescaled state variables $\hat{c}_{s,i}$, $\hat{c}_i$, $\hat{\phi}_i$, and $\hat{\psi}_i$, the formula for the reaction rate $j_i$ is also rewritten as follows:
\[
    j_i = 
    \begin{cases}
        2 kc_{s,i,\text{max}}c_{\text{ref}}^{0.5} \left(1 - \hat{c}_{s,i,\mathrm{surf}}\right)^{0.5} \hat{c}_{s,i,\mathrm{surf}}^{0.5} \hat{c}_i^{0.5} \sinh\left(\dfrac{F}{2RT}\eta_i\right)&\text{for}\ i=p,n\\
        0&\text{for}\ i=s
    \end{cases}
\]
where \(\eta_i\) is given as:
\[
    \eta_i = \psi_{\text{ref}}\hat{\psi}_i - \phi_{\text{ref}}\hat{\phi}_i - U_i\left(\hat{c}_{s,i}\right).
\]
Throughout this paper, we will use $c_{\mathrm{ref}}=10^3$, and $\phi_{\mathrm{ref}} = \psi_{\mathrm{ref}}=1$. In all the following sections, we will consider only the rescaled form of the P2D model, obtained via the above non-dimensionalization. For simplicity, we omit the $\hat{\bullet}$ notation, and all variables will be referred to in their rescaled forms.

\subsubsection*{Initial and boundary conditions}
As shown in the PINN loss \eqref{eq-pinn-loss2}, PINNs generally enforce boundary and initial conditions on the neural network through penalty-type loss functions. This approach, commonly referred to as the soft constraint method or penalty method, has been effective in many cases. However, the imbalance between the PDE residual loss and the penalty loss can lead to training instability and failure, even for relatively simple cases \cite{wong2022learning}.

To address this issue, an alternative approach is to redesign the neural network architecture so that it inherently satisfies the boundary and initial conditions, eliminating the need for additional penalty loss terms. This method, commonly known as the hard constraint method, has been widely applied in PINN applications \cite{lu2021physics}. Adopting this hard-constraint approach, we reconstruct the network architectures in such a way that it automatically satisfies some of the boundary and initial conditions, as follows:

\noindent{\bf The liquid concentration} We use the following structure of network to approximate the liquid concentration $c_i$:
\[
    \widetilde{c}_i(t, x; \theta) = c_i^0( x)\exp\left(t \cdot \mathrm{MLP}(t, x; \theta)\right),\qquad \text{for}\ i=p,s,n,
\]
which directly enforce that the approximate solution $\widetilde{c}(t, x; \theta)$ satisfies the initial condition:
\begin{align*}
    \widetilde{c}_i(0, x; \theta)=c_i^0( x),\qquad \text{for}\ i=p,s,n.
\end{align*}
{\bf The solid concentration} For the solid concentration $c_{s,p}$ in the positive electrode, we use
\begin{align*}
    \widetilde{c}_{s,p}(t, r, x; \theta) &= 1 - \exp\left(-\left(\sqrt{\log\left(\frac{1}{1 - c_{s,p}^0(r, x)}\right)} + t \cdot \mathrm{MLP}(t, r, x; \theta)\right)^2\right),
\end{align*}
and for the solid concentration in the negative electrode,
\begin{align*}
    \widetilde{c}_{s,n}(t, r, x; \theta) &= \exp\left(-\left(\sqrt{\log\left(\frac{1}{c_{s, n}^0(r, x)}\right)} + t \cdot \mathrm{MLP}(t, r, x; \theta)\right)^2\right).
\end{align*}
From the construction of approximate solutions, it follows that
\begin{align*}
    \widetilde{c}_{s,i}(0, r, x; \theta)=c_{s,p}^0(x)=c_{s,i}^0(r,x)\qquad \text{for}\ i=p,n.
\end{align*}
{\bf The solid potential} For the solid potential $\psi_n$ in the negative electrode, we utilize
\[
    \widetilde{\psi}_n(t, x; \theta) = (1 - x) \cdot \mathrm{MLP}(t, x; \theta),
\]
which implies that
\begin{align*}
    \widetilde{\psi}_n(t, 1; \theta) = 0,
\end{align*}
where we note that each $\mathrm{MLP}(\cdot;\theta)$ represents a different fully-connected neural network.

Through this hard-constraint method, the approximate solutions automatically satisfy the initial conditions and Dirichlet boundary conditions without soft constraints. However, for Neumann boundary conditions, we do not apply the hard-constraint method and instead continue using penalty methods, \jm{to be consistent with interfacial conditions which will be discussed below.}

\subsubsection*{Interfacial conditions}\label{subsec-inter}
Unlike boundary or initial conditions, enforcing interfacial conditions through a hard-constraint method is not straightforward, since at least two approximate solutions from different networks are involved at one interface condition. Based on lessons from the additive Schwarz method, it is reasonable to expect that naively imposing interface conditions could lead to reduced information exchange and potentially slower convergence due to limited communication between the individual neural networks.

To explore how PINNs handle complex interface conditions, we tested a  simplified toy example to represent the geometry of a lithium-ion battery:
\begin{equation}\label{eq-toy-interface-z}
\left\{
    \begin{array}{ll}
        u_x = 0, \quad \phi_x = 0 & \quad x = 0, \\[8pt]
        u_t = \dfrac{4}{\pi} u_{xx} - \dfrac{\pi^2}{4} \phi + \cos\left(\dfrac{\pi}{2} x \right) \cos(t), \quad \phi_{xx} = u & \quad x \in (0,1), \\[8pt]
        u = v, \quad 4u_x = v_x, \quad \phi = 0 & \quad x = 1, \\[8pt]
        v_t = \dfrac{1}{4\pi^2} v_{xx} - \sin(2\pi x) \left(\sin(t) + \cos(t)\right) & \quad x \in (1,2), \\[8pt]
        v = w, \quad v_x = 2w_x, \quad \psi = 0 & \quad x = 2, \\[8pt]
        w_t = \dfrac{1}{\pi^2} w_{xx} - \psi - \sin(\pi x) \cos(t), \quad \psi_{xx} = w & \quad x \in (2,3), \\[8pt]
        w = w_\mathrm{sol}, \quad \psi_x = 0 & \quad x = 3.
    \end{array}
\right.
\end{equation}
The exact solution for the equations is given by:
\begin{align}
\begin{split}\label{eq-toy-interface-sol-z}
    u_\mathrm{sol}(x,t) &= \cos\left(\dfrac{\pi}{2}x\right)\sin(t), \\
    v_\mathrm{sol}(x,t) &= -\sin(2\pi x)\sin(t), \\
    w_\mathrm{sol}(x,t) &= -\sin(\pi x) \sin(t), \\
    \phi_\mathrm{sol}(t,x) &= -\dfrac{4}{\pi^2}\cos\left(\dfrac{\pi}{2}x\right)\sin(t), \\
    \psi_\mathrm{sol}(t,x) &= \dfrac{1}{\pi^2}\sin(\pi x)\sin(t). 
\end{split}
\end{align}
This toy example represents a simplified scenario that reflects the essential dynamics of a lithium-ion battery system. It enables us to validate the performance of the domain decomposition method in handling interfacial conditions with discontinuous derivatives. By applying appropriate boundary and interface conditions, this setup captures the behavior of the electrodes and the separator.

\begin{figure}[ht!]
    \centering
\includegraphics[width=1\linewidth]{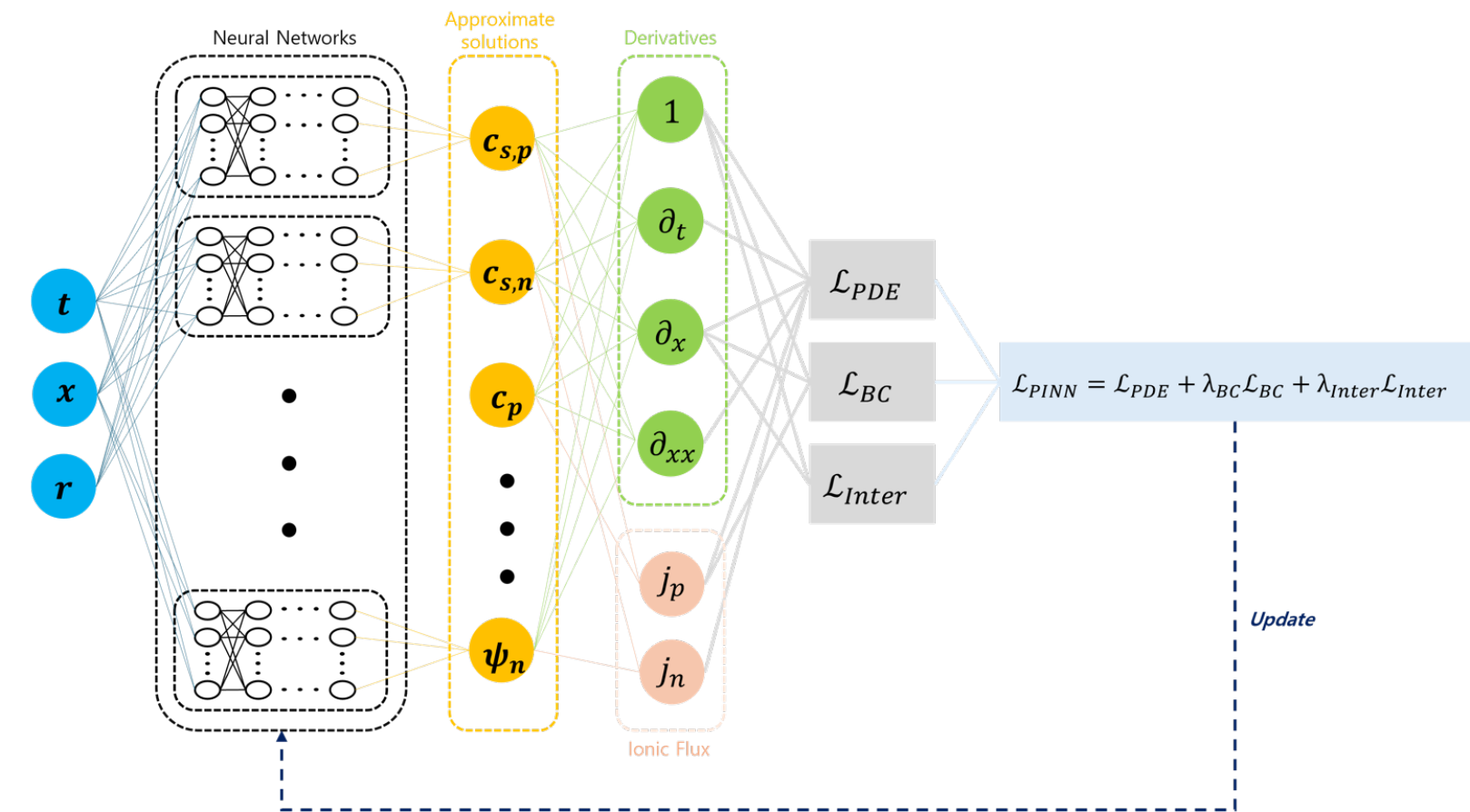}
    \caption{A schematic diagram of the PINN architecture for the P2D model of a Li-ion battery cell.}
    \label{fig-structure1}
\end{figure}

To solve this system using PINNs, we use five neural networks—one for each variable. Interface conditions are incorporated into the PINN loss by adding appropriate penalties.
\begin{figure}[ht!]
    \centering
    \includegraphics[width=1\linewidth]{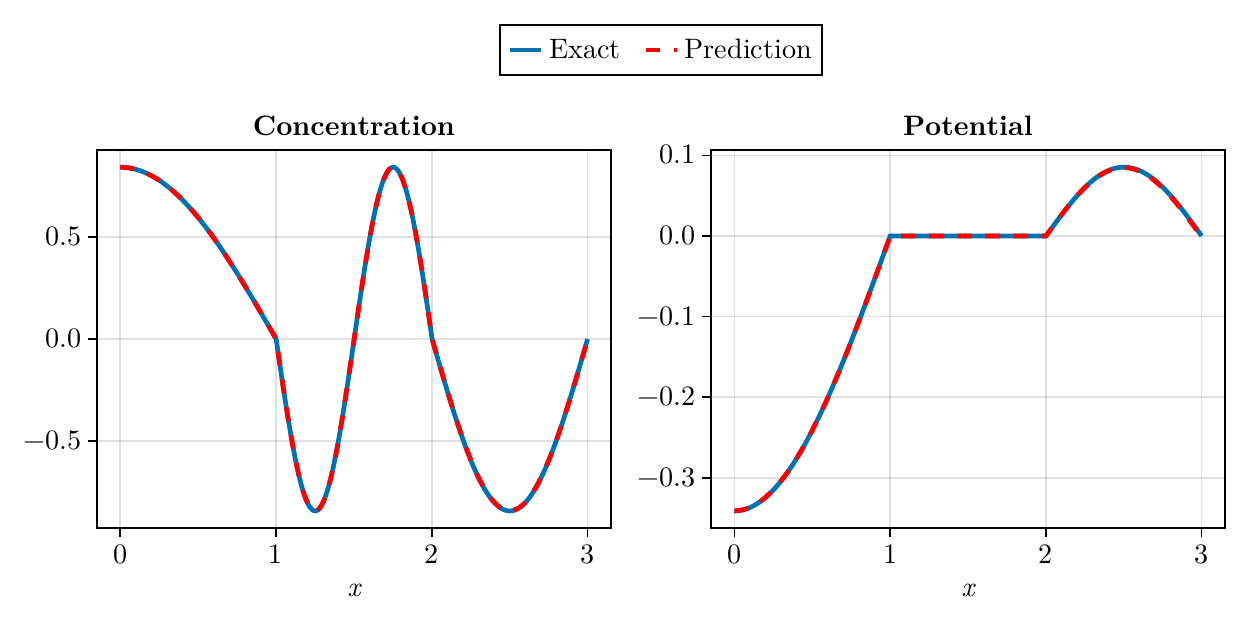}
    \caption{
    Comparison of the PINN result for \eqref{eq-toy-interface-z} to the exact solution \eqref{eq-toy-interface-sol-z}.
    The left figure presents \emph{concentration} variables $u, v$, and $w$.
    The right figure shows \emph{potential} variables $\phi$ and $\psi$.
    Note that potential values are zero in $[1,2]$, to imitate the separator.
    }
    \label{fig-toy-interface}
\end{figure}
\Cref{fig-toy-interface} shows the result of the simulation. Despite the complexity of the system, which includes four boundary conditions, six interface conditions, and five PDEs \eqref{eq-toy-interface-z}, the PINN framework successfully approximates the exact solution with only the penalty method on the interfacial conditions. Through this toy example, we found that a simpler approach—adding soft penalties—is surprisingly sufficient to ensure proper information exchange between the subdomains.
\textcolor{black}{\subsection{Limitations of the conventional approach}
In this section, we presented a vanilla PINN approach for solving the P2D model and then applied existing strategies to address certain issues with that approach. The resulting conventional PINN framework is illustrated in \Cref{fig-structure1}. Notably, all initial conditions are already satisfied by construction through the hard constraint method, making any additional penalty terms for initial conditions unnecessary. However, while these strategies mitigate several difficulties that may arise in solving the P2D model, they remain insufficient for fully resolving complexities of the P2D model. In particular, when using a conventional PINN setup, the initial training loss can be as large as \(O(10^{75})\), leading to severe instability and ultimately causing convergence to fail. In the following section, we focus on the challenges posed by the BV equation, which we believe is a key reason why this conventional approach can fail.}

\section{\textcolor{black}{Challenges related to the Butler--Volmer equation}}\label{subsec-bv}

\begin{figure}[ht]
    \centering
    \includegraphics[width=1\linewidth]{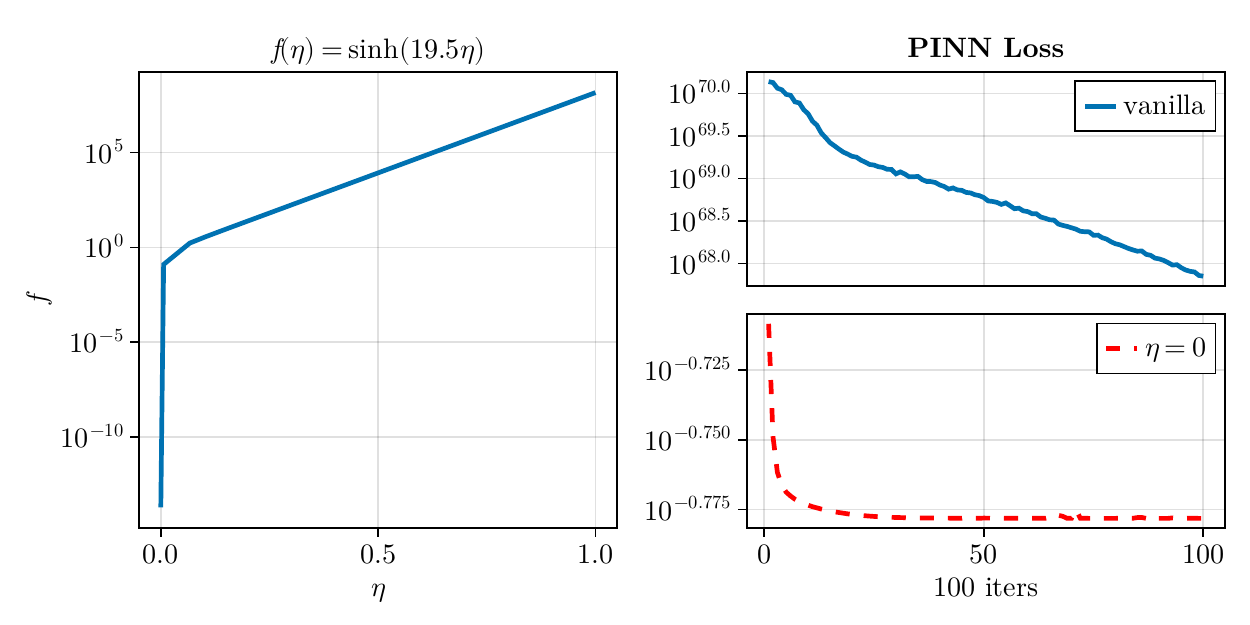}
    \caption{
    The left panel displays the graph of the function \( f(\eta) = \sinh(19.5\eta) \), while the right panel illustrates how errors in \(\eta\) are amplified when evaluating the PINN loss. The solid line represents the baseline, whereas the dashed line corresponds to the scenario where \(\eta = 0\) is manually enforced. These results highlight why the nonlinear BV equation \eqref{BV} is the primary contributor to the large PINN loss values.
    }
    \label{fig-instability}
\end{figure}

\textcolor{black}{In this section, we examine the remaining issues of the conventional PINN approach in \Cref{sec-pinn-p2d}, which lead to an excessively large final loss and ultimately fail to yield accurate solutions. We have identified that these problems stem primarily from the nonlinearity of the BV equation \eqref{BV} in the P2D model. This strong nonlinearity makes the BV equation one of the most challenging parts of the P2D model to address using PINNs.
}
\textcolor{black}{\subsection{Sensitivity of the BV equation with respect to $\eta$}}
We recall the BV equation in its rescaled form:
\begin{align}\label{rescaled-BV}
    j_i \;=\;
    2k\,c_{s,i,\text{max}}\,c_{\text{ref}}^{0.5}
    \,\bigl(1 - c_{s,i,\mathrm{surf}}\bigr)^{0.5}
    \,c_{s,i,\mathrm{surf}}^{0.5}\,c_i^{0.5}
    \,\sinh\!\bigg(\underbrace{\frac{F}{2RT}}_\mathcal{G}\,\eta_i\bigg),
\end{align}
where $\eta_i=\psi_i-\phi_i-U_i$.
Notably, the constant \(\mathcal{G}\) inside the hyperbolic sine function in \eqref{rescaled-BV} is approximately 19.5.
 As shown in the left plot of \Cref{fig-instability}, the graph of \(f(\eta) = \sinh(19.5\eta)\) demonstrates a very fast growth rate as $\eta$ increases. We note that small changes in the approximate solutions $\phi_p$ and $\psi_p$ can lead to large fluctuations in the loss function, exacerbating instability during training.
The impact of this issue is illustrated in the upper right panel of \Cref{fig-instability}, which shows the loss behavior obtained using the conventional PINN approach described earlier. During the early stages of training, the initial loss reaches a magnitude of approximately \(O(10^{75})\). In contrast,  when $\eta$ is manually set to zero, the starting loss value drops to approximately \(O(10^{-1})\), clearly demonstrating the ill-conditioning posed by the BV equation.

\subsection{\textcolor{black}{Discrepancy between loss and accuracy}}
\label{subsec-discrepancy}
\textcolor{black}{A small PINN loss does not always imply that the corresponding approximate solution is sufficiently close to the true solution of the target PDEs. Such discrepancies between loss and accuracy have been reported in the literature \cite{krishnapriyan2021characterizing,wang20222}. In the context of the P2D model, this issue is especially relevant for the BV equation. We recall the rescaled governing equation for the solid potential \(\psi_p\):
\begin{align*}
    \frac{\partial^2 \psi_p}{\partial x^2} 
    \;=\; 
    \frac{L^2 a_p F \, j_p}{\sigma_p^\mathrm{eff} \,\psi_\mathrm{ref}},
    \quad x \in (0,1),
\end{align*}
where the source term on the right-hand side is:
\begin{align}\label{rescaled-BV2}
    \underbrace{\frac{2k\,c_{s,p,\max}\,c_{\text{ref}}^{0.5}\,L^2\,a_p\,F}{
    \sigma^\mathrm{eff}\,\psi_\mathrm{ref}}
    \,\bigl(1 - c_{s,p,\mathrm{surf}}\bigr)^{0.5}
    \,c_{s,p,\mathrm{surf}}^{0.5}\,c_p^{0.5}
    }_{=: \mathcal{H}}
    \;\sinh\Bigl(\tfrac{F}{2RT}\,\eta_p\Bigr),
\end{align}
according to the definition of \(j_p\) in \eqref{BV}. Notably, the coefficient \(\mathcal{H}\) can be much smaller in magnitude than other rescaled variables (on the order of 1). Although the hyperbolic sine function grows exponentially for large inputs, it behaves almost linearly near the origin. In most practical situations, the over-potential \(\eta\) remains small, so \(\sinh\bigl(\frac{F}{2RT}\,\eta_p\bigr) \approx \frac{F}{2RT}\,\eta_p\). Consequently, when \(\mathcal{H}\) is also very small, the right-hand side of \eqref{rescaled-BV2} changes only slightly even if \(\eta_p\) (or \(\psi_p\)) is perturbed.} 

\textcolor{black}{Keeping these points in mind, let us consider a simple example of how inaccurate solutions can still yield small PINN losses. Suppose \(\widetilde{\psi}_p\) differs from the true solution \(\psi_p\) by a constant shift \(C = O(1)\), i.e., \(\widetilde{\psi}_p(x) = \psi_p(x) + C\). Then, \(\widetilde{\psi}_p\) can produce nearly the same small loss as \(\psi_p\), even though \(\widetilde{\psi}_p\) is far from the true solution. The reasons are:
\begin{itemize}
    \item \textbf{PDE residual:} 
    The left-hand side of PDE involves only \(\tfrac{\partial^2\psi_p}{\partial x^2}\) so that a constant shift \(C\) disappears in that side. Also, if \(\mathcal{H}\) is sufficiently small, the source term changes very slightly when shifting \(\psi_p\) by \(C\). Therefore, the shifted one may still lead to a very small PDE residual loss.
    \item \textbf{Neumann boundary conditions:}
    The relevant boundary conditions are Neumann conditions so that a constant shift does not change the boundary loss. 
\end{itemize}
An inaccuracy in \(\psi_p\) can distort other state variables and degrade the solution’s overall accuracy, yet the PINN loss function may fail to detect it. In other words, even if the PINN loss is minimized to a very small value, the approximate solutions may still deviate substantially from the true solution. This discrepancy between loss and accuracy complicates the application of PINNs to the P2D model, underscoring the need for additional strategies to mitigate it.}

In addition, we present the following toy example to illustrate the practical impact of how the small coefficient \(\mathcal{H}\) affects the accuracy of PINN approaches in a simpler setting:
\begin{align}\label{eq-toy-sc}
    \begin{cases}
    \phi_{xx} &= j, \quad x \in (0,1), \\
    \phi_x &= -10^{-4}, \quad x = 0, \\
    \phi &= 0, \quad x = 1,
    \end{cases}
    \qquad \text{and}\qquad
    \begin{cases}
    \psi_{xx} + f &= j, \quad x \in (0,1), \\
    \psi_x &= 0, \quad x \in \{0,1\},
    \end{cases}
\end{align}
where the source term \(j\) is given by
\begin{align*}
    j &= j_0 \left(e^{\psi - \phi + U} - 1\right), 
    \qquad 
    f(x) = (10^{-3} - 10^{-4})(1-x)^8 + 6x - 3, \\
    U(x) &= x^2(x - 1.5) + 4 
    - 10^{-5} (1-x)^{10}
    + \log \Bigl(j_0^{-1} (10^{-3} - 10^{-4})(1-x)^8 + 1 \Bigr),
\end{align*}
and \(j_0\) is a parameter to be chosen. The exact solutions for \(\phi\) and \(\psi\) are:
\[
    \phi(x) = 10^{-5}(1-x)^{10},
    \quad
    \psi(x) = -x^2(x - 1.5) - 4.
\]
Here, \(j_0\) plays a role similar to \(\mathcal{H}\) in the original P2D model.

\begin{figure}[ht]
    \centering
    \includegraphics[width=\linewidth]{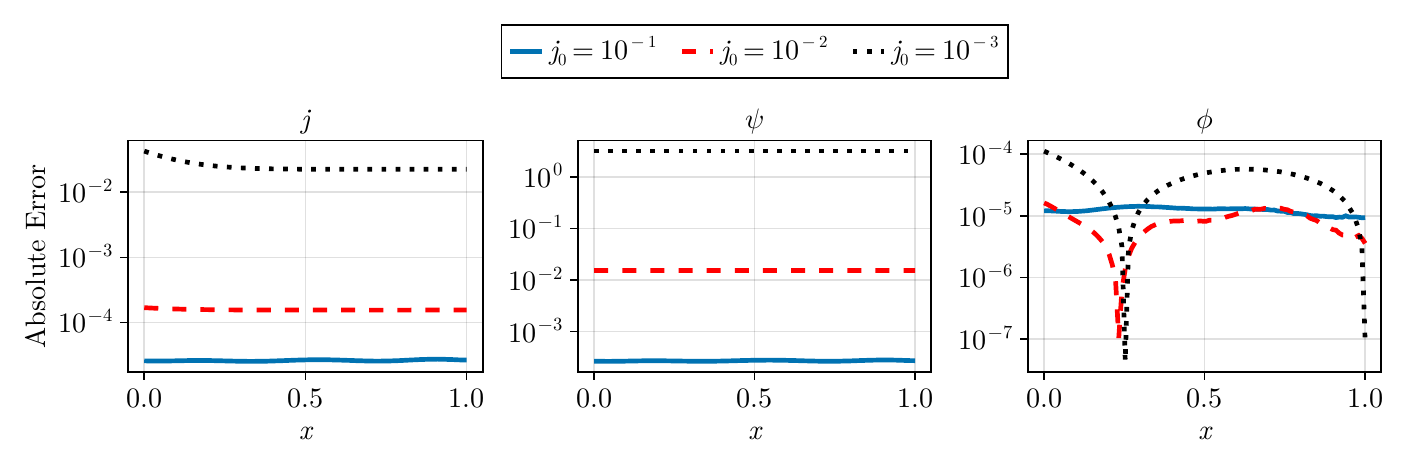}
    \caption{PINN results for the toy model \eqref{eq-toy-sc} with \(j_0 \in \{10^{-1}, 10^{-2}, 10^{-3}\}\). The vertical axes represent absolute error.}
    \label{fig-toy-sc2}
\end{figure}

Let us now consider a shifted function \(\widetilde{\psi} = \psi + C\). Using a Taylor expansion, we have:
\[
    e^{\widetilde{\psi} - \phi + U} - 1 
    \;\approx\;
    e^{\psi - \phi + U} - 1 
    + C\,e^{\psi - \phi + U}
    \;=\;
    \frac{j}{j_0} + C\,e^{\psi - \phi + U}.
\]
Let \(\widetilde{j} = j_0 \bigl(e^{\widetilde{\psi} - \phi + U} - 1\bigr)\). Then
\[
   \widetilde{j}\approx j + C\,j_0\,e^{\psi - \phi + U}.
\]
Since 
\[
    e^{\psi - \phi + U}
    \;=\;
    j_0^{-1} (10^{-3} - 10^{-4})(1-x)^8 + 1
    \;\leq\; j_0^{-1}(10^{-3}-10^{-4})+1,
\]
the shifted function \(\widetilde{\psi}\) alters the source term \(j\) by approximately \(j_0 C + 10^{-3}C\). Hence, the PDE residual of \(\widetilde{\psi}\) becomes
\begin{align*}
    |\widetilde{\psi}_{xx} + f - \widetilde{j}| 
    &= |j - \widetilde{j}|
    \quad(\text{since } \widetilde{\psi}_{xx}=\psi_{xx}),\\
    &= O\bigl(j_0 C + 10^{-3}C\bigr),
\end{align*}
while the boundary loss remains zero (the boundary conditions for \(\psi\) are Neumann). According to this observation, when \(j_0 = 10^{-1}\), the PDE residual loss from \(\widetilde{\psi}\) is on the order of \(O(10^{-2})\). However, if \(j_0\) is much smaller (e.g., \(10^{-3}\)), the PDE residual would be below \(O(10^{-6})\). In typical PINN training, it is generally infeasible to reduce the loss function exactly to zero; the loss typically converges to about \(10^{-4}\) or \(10^{-5}\). Therefore, if the PINN loss by \(\widetilde{\psi}\) is smaller than the achieved loss threshold, the PINN may be unable to distinguish between the true solution \(\psi\) and the shifted function \(\widetilde{\psi}\). In fact, the experimental results in \Cref{fig-toy-sc2} confirm our observation. Specifically, for a relatively large \(j_0 = 10^{-1}\), the PINN provides the correct solution easily, but for a smaller \(j_0 = 10^{-3}\), the trained solution is inaccurate. This confirms that the smaller \(j_0\) is, the more PINNs struggle to distinguish between the true solution and a shifted one. 

This toy example illustrates a key challenge related to the smallness of $\mathcal{H}$ of the BV equation, which hinder the PINN’s ability to accurately capture the true solution. 
\begin{remark}
It is worth highlighting the study by Hassanaly et al.\ \cite{hassanaly2023pinn1}, which encountered similar issues when training PINNs for the single-particle model (SPM) under a nonlinear BV equation. Their solution involved a two-step training approach—first training on a linearized BV equation, then transferring the learned parameters to the nonlinear model for initialization. While innovative, that method introduces extra hyperparameters requiring careful tuning, and their results are not extensively detailed. In contrast, we focus on developing an end-to-end training strategy that avoids these additional complexities, offering a simpler and more robust alternative for handling the nonlinear BV equation in the P2D model.
\end{remark}

\section{Strategies for the nonlinear Butler--Volmer equation}
\label{sec-methods}
As discussed in the previous sections, the highly nonlinear nature of the BV equation can cause significant difficulties in applying PINNs to solve the P2D model. In this section, we propose two strategies aimed at mitigating these challenges: the introduction of a bypassing term to stabilize the training process and the addition of secondary conservation laws to improve the accuracy of solutions.

\subsection{Bypassing term}
\label{subsec-bypassing-term}
The instability caused by the exponential growth of the $\sinh$ term in the BV equation, particularly when applied to PINNs, can be attributed to the sensitivity of the overpotential \(\eta\) to changes in the network parameters. \textcolor{black}{In fact, the PINN approach presented in \Cref{sec-pinn-p2d} approximates the overpotential $\widetilde{\eta}$ using multiple approximate solutions $\widetilde{\psi}$, $\widetilde{\phi}$, $
\widetilde{c}_s$, according to the definition of overpotential $\psi-\phi-U(c_{s,\mathrm{surf}}/c_{s,\mathrm{max}})$. In this way, $\widetilde{\eta}$ depends on several networks, making it not straightforward to initialize $\widetilde{\eta}$ at a small value. Consequently, the approximate ionic flux $\widetilde{j}_i$ become highly sensitive to errors in any of the associated approximations, which results in an instable training. In addition, because each network is also involved in different types of losses terms, $\widetilde{\eta}$ can be affected by various parts of the overall loss function, leading to difficulties in maintaining a proper approximation.} For this, we propose a novel approach to reduce the sensitivity of \(\eta\) during the optimization process, making it possible to handle the fully non-linear BV equation.

Specifically, we introduce a neural network, \(\beta_\theta\), which replaces the entire term inside the hyperbolic sine function in the BV equation. In this approach, the approximate reaction rate $\widetilde{j}_i$ in the PINN loss is computed as  
\begin{equation}\label{eq-zeta}
    \widetilde{j}_i = 2 k c_{s,i,\mathrm{max}}c_{\mathrm{ref}}^{0.5}\left(1 - \widetilde{c}_{s,i,\mathrm{surf}}\right)^{0.5} \widetilde{c}_{s,i,\mathrm{surf}}^{0.5} \widetilde{c}_i^{0.5} \sinh\left(\beta_\theta\right)
\end{equation}
instead of the formulation in \eqref{eq-pinn-j}. Then, an additional penalty term $\mathcal{L}_{\beta}$ is incorporated into the PINN loss to ensure  $\beta_\theta$ accurately approximates $(F/2RT)\eta$: 
\begin{align}\label{bypassing_loss}
    \mathcal{L}_{\beta}:=\int_\Omega \Bigl(\beta_\theta - \dfrac{F}{2RT}\widetilde{\eta}\Bigr)^2\,dx.
\end{align}
We refer to this strategy as \emph{bypassing}. \textcolor{black}{By assigning a dedicated single network to approximate the overpotential \(\eta\), we isolate the approximation inside the hyperbolic sine function from the other networks and prevent large interference among several types of loss terms. Furthermore, this design facilitates initializing $\beta_\theta$ at an $O(1)$ scale, which helps stabilize the early stages of training. Altogether, these mitigate the risk of exploding loss values.}

\begin{figure}[ht]
    \centering
    \includegraphics[width=1\linewidth]{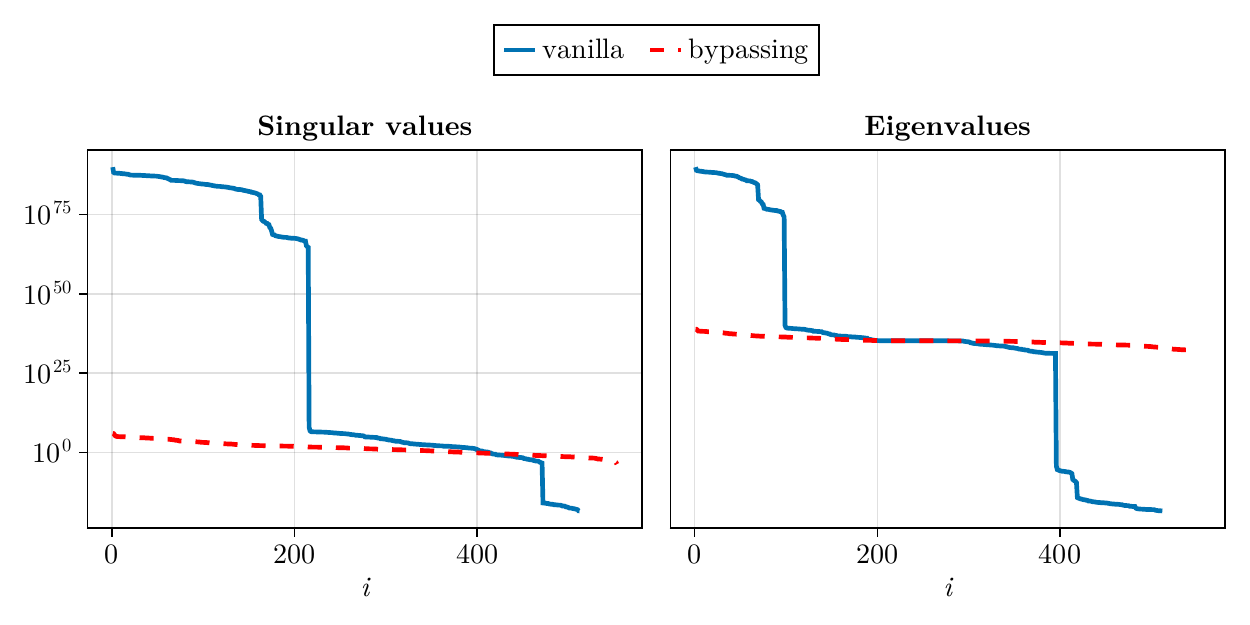}
    \caption{
    {\bf Distribution of singular values and eigenvalues of the Hessian matrix.}
    The solid line represents the distribution without the bypassing term, while the dashed line represents the distribution with the bypassing term. The introduction of the bypassing term significantly narrows the spectrum, indicating improved conditioning. The condition numbers were approximately \(2.017 \cdot 10^{108}\) without the bypassing term and \(4.599 \cdot 10^9\) with the bypassing term.}
    \label{fig-singular-eigen}
\end{figure}

In \cite{wang2021understanding}, Wang et al.\ observed that an imbalance between the gradients of the PDE loss (\(\nabla_\theta \mathcal{L}_\mathrm{PDE}\)) and the boundary condition loss (\(\nabla_\theta \mathcal{L}_\mathrm{BC}\)) can lead to a wide range of eigenvalues in the Hessian matrix, resulting in ill-conditioned training dynamics. This emphasizes the importance of maintaining a balanced gradient distribution for effective training. From this point of view, we can verify the effectiveness of the bypassing strategy. By conducting an eigenvalue analysis, we observed a significant reduction in the condition number after introducing the bypassing term. Although memory constraints required us to perform this analysis with a smaller neural network compared to the one used for the full P2D model in \Cref{subsec-p2d-result}, the results still sufficiently demonstrate the effectiveness of the proposed strategy. The results are visualized in \Cref{fig-singular-eigen}.

\begin{remark}
    Our bypassing strategy is inspired by observations from previous studies. Han et al. \cite{han2021fast} successfully solved the P2D model using a finite difference method with backward Euler time stepping. Notably, their approach involved treating \(\eta\) and \(j\) as independent unknowns, despite these quantities being defined by algebraic equations. We hypothesize that this choice was made to address stability issues inherent to the BV equation. Building on their insight, we adopted a similar concept by introducing a neural network, \(\beta_\theta\), to approximate a scaled version of \(\eta\).
\end{remark}

\subsection{Secondary conservation laws}\label{subsec-sc}
As mentioned in \Cref{subsec-bv}, due to the small scaling factor $\mathcal{H}$ in the BV term \eqref{rescaled-BV}, the conventional PINN loss struggles to capture the true battery state, particularly for $\psi_p$. A function that is quite far from the true solution can still result in a very small loss, and further reducing this loss would require significant computational resources because the corresponding loss value is already small. In other words, minimizing the PINN loss presented in \Cref{sec-pinn-p2d} is insufficient for accurately approximating the true solution.

\begin{figure}[ht!]
    \centering
    \includegraphics[width=1\linewidth]{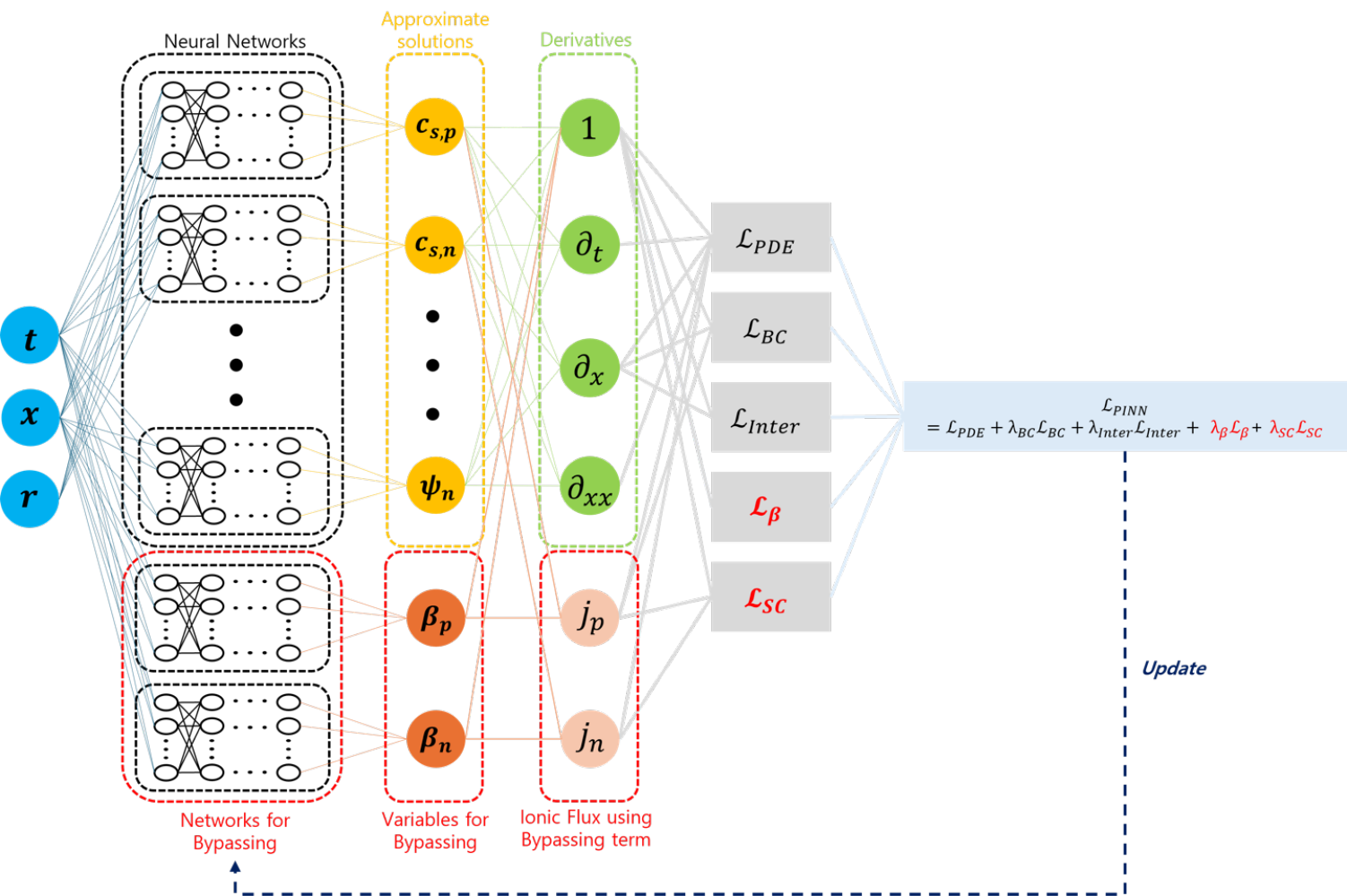}
    \caption{A schematic diagram of our proposed PINN architecture for the P2D model of a Li-ion battery cell.}
    \label{fig-structure2}
\end{figure}

\textcolor{black}{To mitigate this discrepancy between loss and accuracy}, we introduce additional constraints based on secondary conservation laws to more accurately approximate the reaction rates $\widetilde{j}_i$, which, in turn, guide the approximate solutions toward the true solution. Specifically, we enforce the following secondary conservation laws: \begin{equation}
    \frac{L a_p F}{I_\mathrm{app}} \int_0^{\hat{L}_p} j_p \, dx = 1, \quad \text{and} \quad \frac{L a_n F}{I_\mathrm{app}} \int_{\hat{L}_p + \hat{L}_s}^{1} j_n \, dx = -1,
\end{equation}
which are derived by integrating \eqref{eq-phi-pde-nd}. These constraints help ensure the correct computation of the reaction rates, which in turn guides the approximation of $\widetilde{\psi}_p$ and other key variables in the P2D model. Our approach, which incorporates the bypassing terms and secondary conservation laws, is illustrated in \Cref{fig-structure2}.

\begin{figure}[ht]
\centering
    \includegraphics[width=1\linewidth]{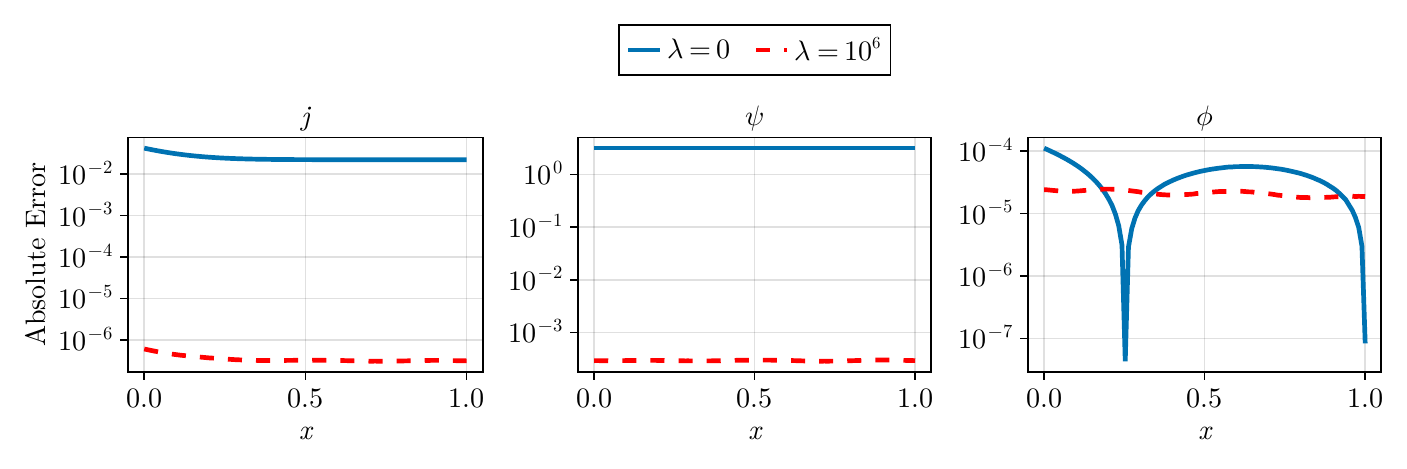} \caption{Results for \(j_0=10^{-3}\) after adding $\lambda L_\mathrm{SC}(\theta)$ to the PINN loss. Solid lines represent the baseline, whereas dotted lines show the result of adding the secondary conservation law. For $j$ and $\psi$, the error magnitudes are reduced by several orders of magnitude.}
    \label{fig-toy-sc}
\end{figure}

For the toy example presented in \Cref{subsec-bv}, we can see that this strategy helps prevent the network from converging to a shifted or incorrect solution. For the toy example, the corresponding secondary conservation can be expressed as: \begin{align}\label{eq-sc2} \int_0^1j\ dx=10^{-4}. 
\end{align} As previously discussed, using a typical PINN loss alone is not sufficient for solving the toy example effectively, especially when dealing with the small factor $j_0$. However, by incorporating the additional constraints \eqref{eq-sc2} into the PINN loss function, the PINNs can predict the solution more accurately, as shown in \Cref{fig-toy-sc}.

As demonstrated in the toy example, by combining secondary conservation laws with the PINN framework, we can mitigate the challenges posed by the BV equation and improve the network's ability to capture the true solution with greater accuracy, despite the computational challenges.

\subsection{Ablation study}
\label{subsec-p2d-ablation}

We performed an ablation study to evaluate the individual and combined effectiveness of the methods proposed in this section. The results are summarized in Table~\ref{tab-ablation}.

\begin{table}[ht]
    \centering
    \begin{tabular}{c|c|c|c|c|c|c}
    \hline
    Methods & $\psi$ & $c_s$ & $\phi$ & $c$ & time(s) & final loss \\
    \hline\hline
    Baseline & 8.69e-1 & 7.60e-1 & 5.18e-1 & 1.89e-1 & 31049 & 1.63e+48\\
    \hline
    B & 2.24e-1 & 8.96e-1 & 8.88e-1 & 2.11e-1 & \textbf{4729} & 7.32e-8\\
    \hline
    SC & 8.69e-1 & 7.61e-1 & 5.17e-1 & 1.89e-1 & 50578 & 1.63e+48\\
    \hline
    B + SC & \textbf{4.84e-3} & \textbf{1.31e-2} & \textbf{2.31e-2} & \textbf{6.51e-3} & 6352 & 9.15e-6 \\
    \hline
    \end{tabular}
    \caption{
    Ablation study results showing relative \(L^2\) errors.  
    \textbf{B}: Using the bypassing term to approximate the stiffest component of the BV equation.  
    \textbf{SC}: Using the secondary conservation law to enhance solution accuracy by preventing convergence to trivial solutions.  
    \textbf{B + SC}: Using both strategies simultaneously.
    }
    \label{tab-ablation}
\end{table}

\noindent
The Baseline configuration represents the simplest implementation of the PINN framework presented in \Cref{sec-pinn-p2d}, without both the bypassing term and the secondary conservation law. The B configuration includes the bypassing term to address the stiffness of the BV equation, whereas the SC configuration incorporates the secondary conservation law into the PINN loss to mitigate the discrepancy between loss and accuracy.

\subsubsection*{\jm{Necessity of bypassing terms}}
\jm{The Baseline and SC configurations both exhibit extremely large losses (on the order of \(10^{48}\)) and require over 8 hours of training, as neither addresses the high sensitivity of the BV equation. This underscores the need to mitigate the ill-conditioning posed by the nonlinear BV term. By employing the bypassing method, we successfully reduce the PINN loss to \(7.32\times10^{-8}\), indicating that the instability by the BV equation has been effectively resolved. The bypassing method remains effective when combined with the secondary conservation law, achieving a still low loss value of \(9.15\times10^{-6}\).}

\subsubsection*{\jm{Necessity of secondary conservation laws}}
\jm{Although the bypassing term ensures faster training and a small loss, the accuracy achieved with the bypassing method alone is still insufficient. In other words, achieving a small value for the PINN loss (\ref{eq-pinn-loss2}) with bypassing loss (\ref{bypassing_loss}) does not guarantee accuracy. As noted in \Cref{subsec-sc}, the secondary conservation laws are required to resolve this discrepancy. Indeed, once a sufficiently small loss is achieved through the bypassing method, the secondary conservation plays an important role in resolving the discrepancy between loss and accuracy, improving accuracy to the order of $10^{-2}$ or lower.\\}

Overall, our ablation study shows that neither method is redundant; both the bypassing term and the secondary conservation law are essential for an efficient and accurate simulation of the P2D model. This result underscores the robustness of our proposed framework in overcoming the challenges posed by nonlinear BV equations in PINNs.

\section{Numerical experiments}
\label{sec-results}
We performed numerical experiments for both forward and inverse problems of the P2D model to evaluate the effectiveness of our proposed methods. In our numerical experiments, we used a neural network architecture with a width of 32 and a depth of 4. The optimization process followed a two-step approach, combining ADAM \cite{kingma2014adam} and L-BFGS \cite{liu1989limited} optimizers, implemented in the Optax package \cite{deepmind2020jax} and the JAXopt package \cite{jaxopt2021}, respectively. Collocation points were dynamically sampled for ADAM but kept fixed during L-BFGS, using $2^{14}$ points at each step. The ADAM optimizer employed a cosine decay learning rate schedule, starting from $10^{-4}$, and the L-BFGS history size was set to 50. Our network utilized the sine activation function, initialized with the SIREN scheme \cite{sitzmann2020implicit}. \jm{For reference solution, we used $\Delta t = 1$ for time-stepping up to the final time $\tau = 3500$, and $N_x = 50$ for spatial discretization.}
Computations were carried out on an Nvidia A100 GPU, and the framework JAX \cite{jax2018github} was used for automatic differentiation, while visualizations were created using Makie.jl \cite{Makie2021}.

\subsection{Forward problem}\label{subsec-p2d-result}
\begin{figure}[ht]
    \centering
    \includegraphics[width=1\linewidth]{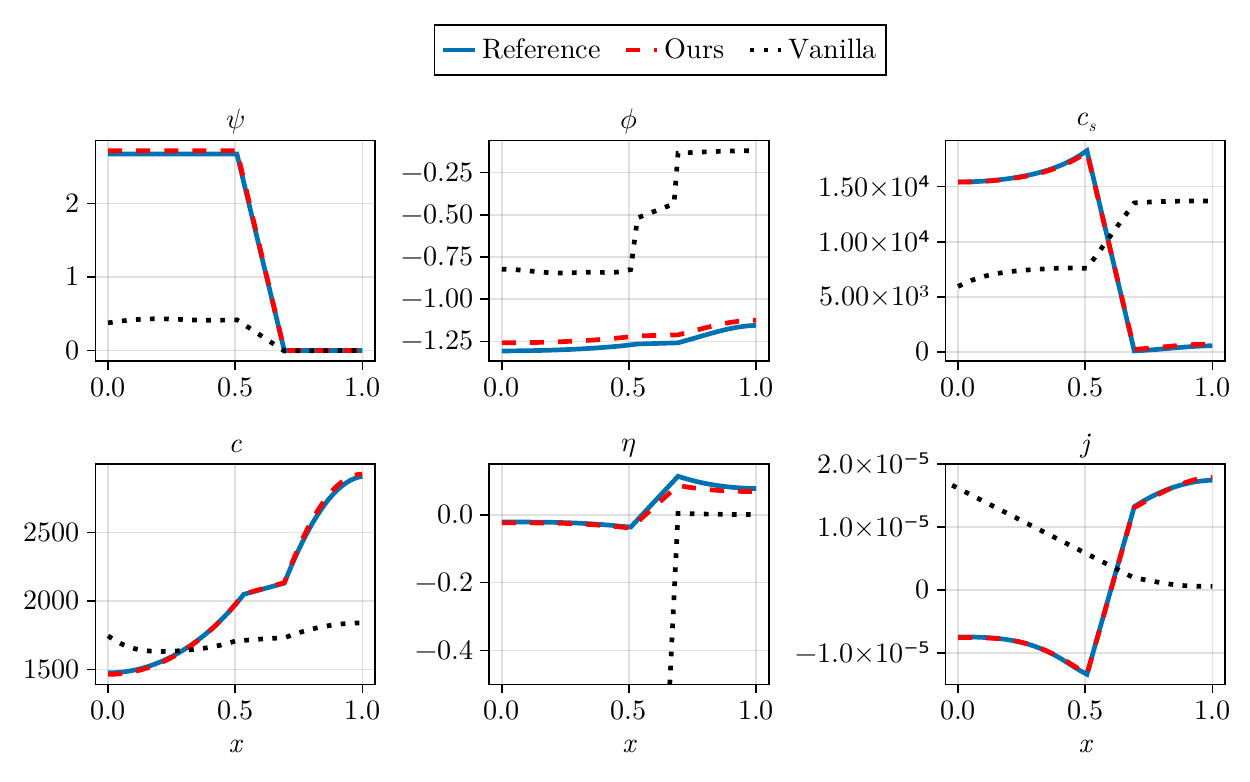}
    \caption{Battery state variables over the spatial domain, evaluated at \(t = 3500\). Our proposed method closely matches the reference solution, whereas vanilla PINNs fail to provide accurate predictions.}
    \label{fig-over-x}
\end{figure}

\begin{figure}[ht]
    \centering
    \includegraphics[width=1\linewidth]{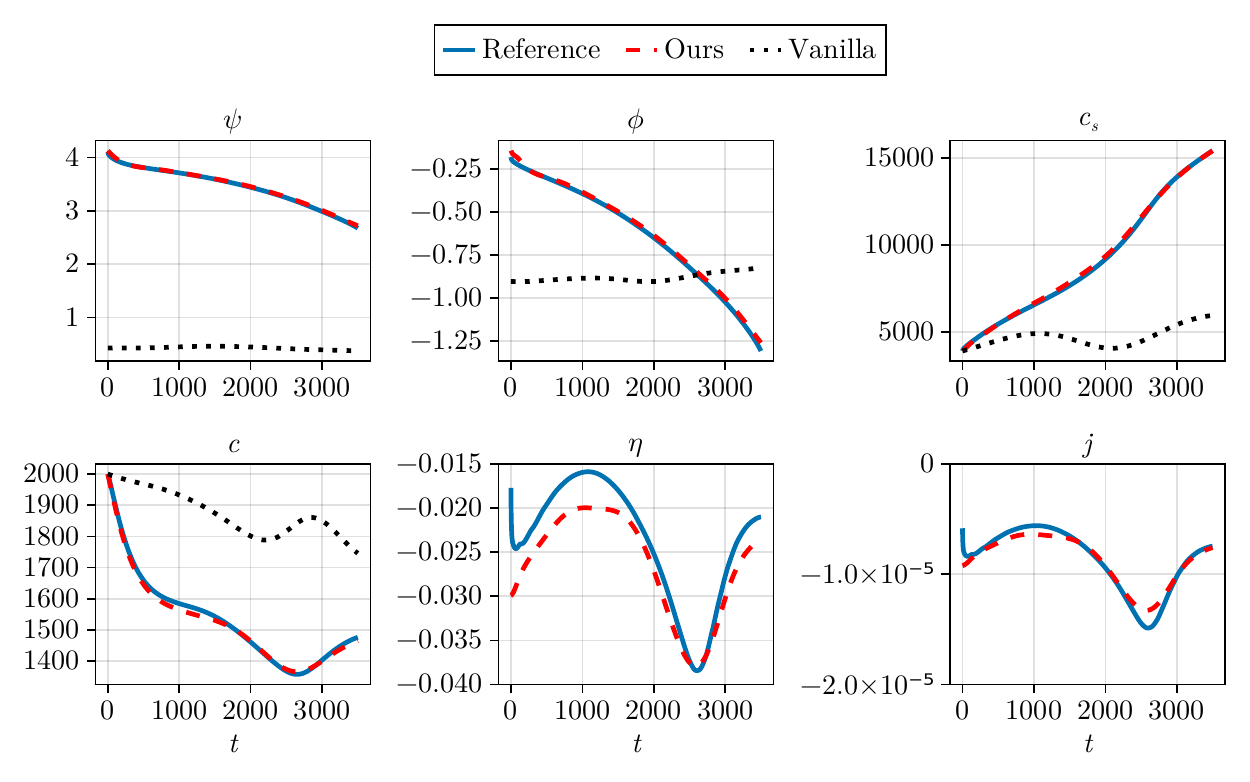}
    \caption{Battery state variables over time, evaluated at the left end of the battery. Our method demonstrates excellent agreement with the reference solution.}
    \label{fig-over-t}
\end{figure}

We present the PINN simulation results for the forward problem of the P2D model, demonstrating the effectiveness of our proposed methods in addressing the challenges posed by the fully nonlinear BV equation \eqref{BV}. The simulations were performed using 100k iterations of the Adam optimizer, followed by 100k iterations of the L-BFGS optimizer. 
\Cref{fig-over-x,fig-over-t} illustrate the spatial distribution and temporal evolution of key battery state variables, evaluated at \(t = 3500\) and \(x = 0\), respectively. Unlike vanilla PINNs, which fail to produce meaningful predictions and deviate significantly from the reference solution, our proposed approach delivers accurate results. \jm{\ref{app-more-experiments} includes results for additional experiments with different physical parameters.} This close agreement between our PINN solution and the reference solution not only validates the robustness of our computational strategies but also highlights their predictive accuracy over long time horizons. 


\begin{figure}[ht!]
    \centering
    \includegraphics[width=1\linewidth]{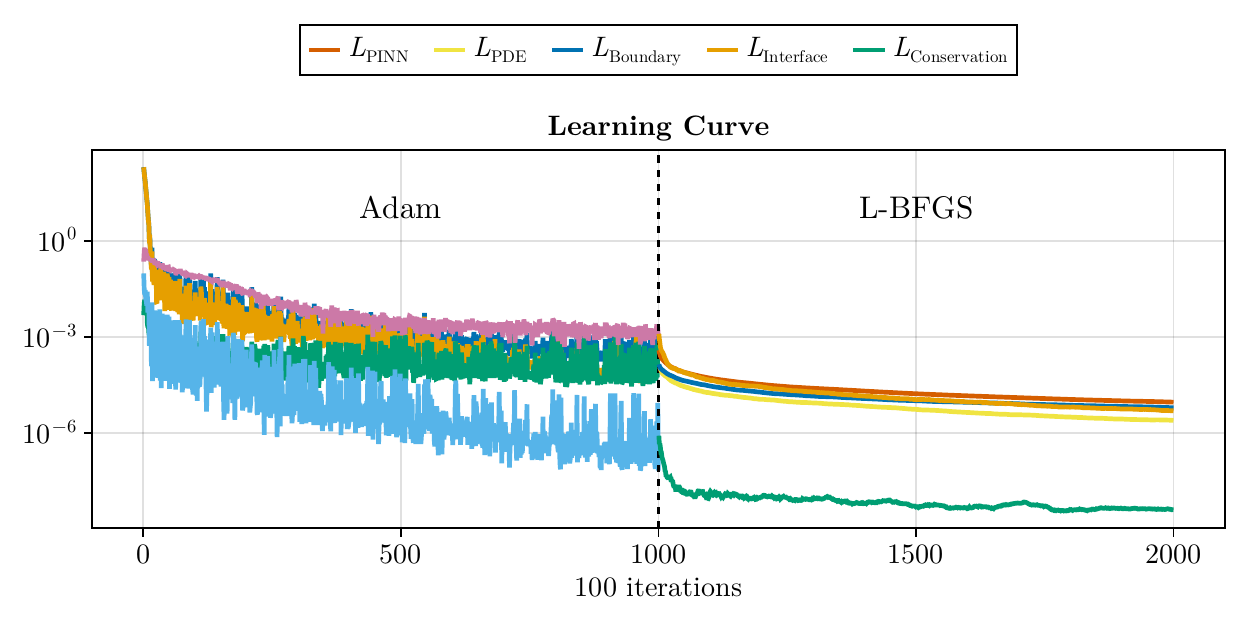}
    \caption{Learning curves. The left panel shows the loss and error trajectories during optimization with the Adam optimizer, while the right panel shows the corresponding trajectories for the L-BFGS optimizer. Please refer to the online version for colored curves.}
    \label{fig-loss-traj}
\end{figure}

The learning curves for the PINN optimization are shown in \Cref{fig-loss-traj}. The left panel illustrates the convergence behavior of the Adam optimizer, while the right panel depicts the results for L-BFGS. Our approach achieves a relative error on the order of \(10^{-2}\), demonstrating sublinear convergence, which is characteristic of PINN optimization problems. 
To the best of our knowledge, this is the first successful PINN simulation of a Li-ion battery incorporating the fully nonlinear BV equation \eqref{BV}. The ability of our method to accurately predict battery behavior underscores the high predictive power and reliability of our computational strategy. These results provide a strong foundation for future research into advanced deep-learning methods for complex battery simulations.

\subsection{Inverse problem}\label{subsec-p2d-inverse}
One advantage of the PINN framework is its ability to estimate model parameters with minimal implementation modifications. Given observational data \(\{u(x_i)\}_{i=1}^N\), we incorporate a data fitting term into the PINN loss \(\mathcal{L}_\text{PINN}\) \eqref{eq-pinn-loss} and treat the unknown parameters \(\alpha\) as learnable parameters. This leads to the following loss function for the inverse problem:
\[
\mathcal{L}(\theta, \alpha) = \int_\Omega \left\lvert\mathcal{N_\alpha}[u_\theta](x) - f(x)\right\rvert_2^2 \, dx + \lambda \int_{\partial \Omega} \left\lvert \mathcal{B_\alpha}[u_\theta](s) - g(s) \right\rvert_2^2 \, d\sigma(s) + \lambda_\mathrm{data} \sum_{i=1}^N \left(u_\theta(x_i) - u(x_i)\right)^2.
\]
The minimizer \(\left(u_{\theta^\star}, \alpha^\star\right)\) of this loss function will yield a solution \((u_{\theta^\star}, \alpha^\star)\) that satisfies the governing PDE and boundary conditions, while also fitting the observed data closely, thus providing a solution to the inverse problem.
With a more reliable method for solving the forward problem now in place, we are prepared to address inverse problems using the reference solution as our data. Our focus is on inverse problems related to battery length, which is considered a crucial factor in the performance of Li-ion batteries \cite{kang2022electrode}.
In the following, we perform experiments on three inverse tasks: estimating the battery length (\Cref{tab-inverse1}), the length ratio (\Cref{tab-inverse2}), and both battery length and length ratio simultaneously (\Cref{tab-inverse3}). These tasks are evaluated under two scenarios: the first uses noise-free observational data, while the second introduces 5\% Gaussian noise into the observational data.
Our proposed methodology demonstrates strong predictive performance across all scenarios. Notably, even in the presence of noisy observational data, the predictions remain comparable to those obtained with clean data. This robustness to noise underscores the reliability and effectiveness of our approach in solving inverse problems, even under challenging conditions.
\\

\noindent
{\bf Battery length}
The first example involves estimating the total battery length, \(L\), while maintaining a fixed ratio of \(L_p: L_s: L_n = 1.74: 0.52: 1.0\). Although this problem involves only a single parameter, it is nontrivial due to its impact on 11 coefficients in the P2D model after non-dimensionalization, particularly those associated with spatial partial derivatives along \(x\). To parameterize \(L\), we expressed it as \(\hat{L} \cdot 10^{-4}\) with an initial value of \(\hat{L} = 1\). After 150,000 iterations of the Adam optimizer, we obtained \(\hat{L} = 3.21\), corresponding to a relative error of 1.51\% compared to the true value \(L = 3.26\). The entire computation was completed in approximately 30 minutes.

\begin{table}[ht]
\centering
\begin{tabular}{c|c|c}
\hline
\textbf{Method} & w/o Noise  & w/ Noise\\ \hline\hline
Baseline        & 69.53 & 69.53\\ \hline
B               & 44.71 & 44.85\\ \hline
SC              & 69.59 & 69.59\\ \hline
B + SC          & {\bf 1.51}  & {\bf 1.58} \\ \hline
\end{tabular}
\caption{\jm{Relative errors (\%) for the total length (\(L\)) estimation problem.}}
\label{tab-inverse1}
\end{table}

\noindent
{\bf Length ratio}
The second example focuses on estimating the ratio of the battery section lengths, \(L_p: L_s: L_n\), while keeping the total battery length fixed at \(L = 3.26 \cdot 10^{-4}\). This task introduces additional complexity due to the dynamically changing domain lengths during optimization, which limits the effectiveness of the L-BFGS optimizer. To simplify the parameterization, we used two parameters, \(\rho_p\) and \(\rho_n\), since the constraint \(\rho_p + \rho_s + \rho_n = 1\) reduces the problem to two degrees of freedom. The initial values were set to \(\hat{\rho}_p = \hat{\rho}_n = 0.4\). The B + SC method estimated \(\hat{L}_p = \hat{\rho}_p \cdot L = 1.68 \cdot 10^{-4}\) and \(\hat{L}_n = \hat{\rho}_n \cdot L = 0.9536 \cdot 10^{-4}\), achieving relative errors of 3.21\% and 4.64\% for the positive and negative electrode lengths, respectively. The computation was completed in approximately 40 minutes.

\begin{table}[ht]
\centering
\begin{tabular}{c|c|c|c|c}
\hline
\multirow{2}{*}{\textbf{Method}} & \multicolumn{2}{|c|}{w/o Noise} & \multicolumn{2}{|c}{w/ Noise} \\ \cline{2-5}
 & $L_p$ & $L_n$ & $L_p$ & $L_n$ \\ \hline\hline
Baseline        & 26.14 & 4.65  & 26.14 & 4.68 \\ \hline
B               & {\bf 2.58}  & {\bf 4.33}  & {\bf 2.71}  & {\bf 4.30} \\ \hline
SC              & 17.10 & 10.01 & 17.10 & 10.23 \\ \hline
B + SC          & 3.21  & 4.64  & 3.10  & 4.81\\ \hline
\end{tabular}
\caption{\jm{Relative errors (\%) for the length ratio (\(L_p, L_n\)) estimation problem.}}
\label{tab-inverse2}
\end{table}

\noindent
{\bf Battery length and length ratio}
Finally, we extended the problem to simultaneously estimate both the battery length and the length ratio. This combined task represents a significantly more challenging problem due to the increased dimensionality and the inherent coupling between the two parameters. Accurate estimation in such scenarios requires the framework to handle dynamic changes in both domain lengths and parameter dependencies throughout the optimization process. Starting with the same initialization for \(\hat{L}, \hat{\rho}_p\), and \(\hat{\rho}_n\), the B + SC method estimated \(\hat{L} = 2.971\), \(\hat{L}_p = 1.614 \cdot 10^{-4}\), and \(\hat{L}_n = 0.9374 \cdot 10^{-4}\). The relative errors for these estimates were \(8.85 \cdot 10^{-2}\), \(7.23 \cdot 10^{-2}\), and \(6.26 \cdot 10^{-2}\), respectively.

\begin{table}[ht]
\centering
\begin{tabular}{c|c|c|c|c|c|c}
\hline
\multirow{2}{*}{\textbf{Method}} & \multicolumn{3}{|c|}{w/o Noise} & \multicolumn{3}{|c}{w/ Noise} \\ \cline{2-7}
& $L$ & $L_p$ & $L_n$ & $L$ & $L_p$ & $L_n$ \\ \hline\hline
Baseline        & 69.66 & 77.67 & 40.01 & 69.66 & 77.67 & 39.98 \\ \hline
B               & 52.01 & 10.86 & 13.45 & 52.19 & 10.74 & 13.10 \\ \hline
SC              & 69.51 & 77.57 & 39.16 & 69.51 & 77.57 & 39.13 \\ \hline
B + SC          & {\bf 8.85}  & {\bf 7.23} & {\bf 6.26} & {\bf 8.89} & {\bf 7.39} & {\bf 6.19} \\ \hline
\end{tabular}
\caption{\jm{Relative errors (\%) for the total length and length ratio (\(L, L_p, L_n\)) estimation problem.}}
\label{tab-inverse3}
\end{table}

The results throughout 3 different inverse problems demonstrate the robustness of the B + SC method in tackling complex coupled inverse problems.

\subsection{A note on implementation}
The P2D model tightly interconnects battery state variables. For instance, calculating the ionic flux $j$ necessitates computing $\beta$, $c$, and $c_{s, \mathrm{surf}}$ for the BV equation.
While frameworks like JAX-PI \cite{wang2021understanding} offer readable implementations, they can be computationally inefficient due to redundant calculations. As shown in \Cref{subfig-inefficient-code}, a JAX-PI-style implementation repeatedly computes \(j\), hindering its reuse for the potential and surface concentration PDEs.
To optimize computational efficiency, we adopt a strategy that prioritizes minimizing network forward passes. By initially calculating all required quantities collectively and then assembling the PDE residuals, we significantly reduce redundant computations. This approach, illustrated in \Cref{subfig-efficient-code}, is expected to yield significant speedups. Our implementation demonstrates this, achieving roughly 1.5 times the performance of a naive approach (14 iterations per second compared to 9).

\begin{figure}[!ht]
\centering

\begin{subfigure}[b]{1\textwidth}
\centering
\begin{minted}[mathescape, breaklines, frame=single, fontsize=\footnotesize]{python}
import jax

def psi_pde(p2d, net, t, x):
    psi_xx = jax.jacfwd(jax.jacfwd(net.psi, 1), 1)(t, x)
    cs_surf = net.cs(t, 1.0, x)
    c = net.c(t, x)
    beta = net.beta(t, x)
    j = p2d.compute_j(beta, cs_surf, c)
    return psi_xx - (p2d.L**2 * p2d.a * F * j)
...
\end{minted}

\caption{Implementation like JAX-PI.}
\label{subfig-inefficient-code}
\end{subfigure}
\hfill
\begin{subfigure}[b]{1\textwidth}
\centering
\begin{minted}[mathescape, breaklines, frame=single, fontsize=\footnotesize]{python}
def compute_quantities(net, t, x):
    psi_xx = jax.jacfwd(jax.jacfwd(net.psi, 1), 1)(t, x)
    cs_surf = net.cs(t, 1.0, x)
    c = net.c(t, x)
    beta = net.beta(t, x)
    ...
    return {"psi_xx":psi_xx, "cs_surf":cs_surf, "c":c, "beta":beta, ...}

def compute_psi_pde(p2d, psi_xx, j):
    return psi_xx - (p2d.L**2 * p2d.a * F * j)
...

def compute_pdes(p2d, quantities):
    j = p2d.compute_j(quantities["beta"], quantities["cs_surf"], quantities["c"])
    psi_pde = compute_psi_pde(quantities["psi_xx"], j)
    cs_pde = compute_cs_pde(quantities["cs_rr"], quantities["cs_r"], j)
    ...
    return pdes
\end{minted}

\caption{Our implementation}
\label{subfig-efficient-code}
\end{subfigure}

\caption{Python pseudo-code}
\end{figure}

\section{Discussion and conclusion}
\label{sec-discussions}
This study has addressed the challenges of simulating Li-ion batteries using the P2D model in combination with PINNs and has proposed effective strategies to overcome these obstacles. Our methods enable reliable and accurate simulations of the P2D model, incorporating the fully nonlinear BV equation. By developing an enhanced forward solver, we successfully tackled inverse problems related to battery section lengths, achieving remarkable accuracy and computational efficiency. Despite the inherent complexity of these problems, our framework delivers results that hold significant engineering relevance for the design and optimization of Li-ion batteries. These achievements highlight the practical applicability of our approach and its ability to solve inverse problems that might not be infeasible for traditional numerical methods.
However, there is still room for further development. For instance, extending our framework to incorporate multiphysics phenomena such as thermal effects or aging, which are currently excluded from our equations, presents an exciting and challenging avenue for future research. Additionally, building on our forward problem solver to develop an operator network capable of real-time predictions is another promising direction. 
% In conclusion, our study has explored the challenges of simulating Li-ion batteries using the P2D model in combination with PINNs and has proposed effective strategies to address these issues. Our methods facilitate reliable simulations of the P2D model, incorporating the nonlinear Butler-Volmer equation. By utilizing an enhanced forward solver, we have made significant progress in solving inverse problems related to battery section lengths. However, several challenges remain unresolved.
% One of the primary obstacles to the industrial application of PINNs is the lengthy training time. Although PINNs offer a promising approach to solving complex PDEs, their computational demands can be substantial. While our methods significantly improve training efficiency, real-time simulation remains unattainable. Additionally, for inverse problems, our techniques have demonstrated that PINNs can handle noisy observation data; however, practical scenarios often present challenges due to sparse observation data, which is frequently available only in statistical forms. 

% Therefore, further research is crucial to develop more efficient training algorithms and architectures that enhance the competitiveness of PINNs relative to traditional numerical methods, especially for integration with battery management systems.

\section*{Acknowledgements}
The work of M.-S. Lee was supported by Basic Science Research Programs through the National Research Foundation of Korea (NRF) funded by the Ministry of Education (RS-2023-00244475 and RS-2024-00462755). The work of Y. Hong was supported by the Basic Science Research Program through the National Research Foundation of Korea (NRF), funded by the Ministry of Education (NRF-2021R1A2C1093579), the Korean government (MSIT) (RS-2023-00219980), and Hyundai Motor Company. This work was supported by the National Supercomputing Center with supercomputing resources including technical support (KSC-2023-CRE-0334).

\bibliographystyle{elsarticle-num}
\bibliography{library}

\appendix
\section{\jm{Additional experiments}}\label{app-more-experiments}
\jm{In addition to the original 1C discharge rate which corresponds to \(\tau = 3500\) and \(I_\mathrm{app} = -17.5\), we test our methods to two different discharge rates, 0.5C and 2C. Specifically, 0.5C discharge rate is $\tau = 7000$ and $I_\mathrm{app} = -8.75$, and 2C discharge rate corresponds to $\tau = 1550$ and $I_\mathrm{app} = -35$. \Cref{tab-discharge-rates} illustrates our method's ability to simulate different discharge rates.
}
\begin{table}[ht]
    \centering
    \begin{tabular}{c|c|c|c|c}
    \hline
    Discharge rate & \(\psi\) & \(c_s\) & \(\phi\) & \(c\) \\
    \hline\hline
    1C & 4.84e-3 & 1.31e-2 & 2.31e-2 & 6.51e-3 \\
    \hline
    0.5C & 4.87e-3 & 2.43e-2 & 1.28e-2 & 3.58e-3 \\
    \hline 
    2C & 1.65e-2 & 4.69e-2 & 7.15e-2 & 6.02e-3 \\
    \hline
    \end{tabular}
    \caption{\jm{Test results (relative \(L^2\) errors) for different discharge rates.}}
    \label{tab-discharge-rates}
\end{table}

\end{document}